\documentclass[preprint]{JHEP3} 

\JHEPspecialurl{http://jhep.sissa.it/JOURNAL/JHEP3.tar.gz}

\usepackage{epsfig,multicol,bbm}
\usepackage{latexsym} 
\usepackage{amssymb}  

\newcommand\fverb{\setbox\pippobox=\hbox\bgroup\verb}
\newcommand\fverbdo{\egroup\medskip\noindent%
			\fbox{\unhbox\pippobox}\ }
\newcommand\fverbit{\egroup\item[\fbox{\unhbox\pippobox}]}
\newbox\pippobox

\newcommand{\p}{\partial}
\newcommand\eqn[1]{(\ref{#1})}      
\newcommand\Eqn[1]{Eq.~(\ref{#1})}  
\newcommand\Ref[1]{Ref.~\cite{#1}}  
\newcommand\Sec[1]{Sec.~\ref{#1}}  

\newcommand{\beq}{\begin{equation}}
\newcommand{\eeq}{\end{equation}}
\newcommand{\ba}{\begin{array}}
\newcommand{\bea}{\begin{eqnarray}}
\newcommand{\ea}{\end{array}}
\newcommand{\eea}{\end{eqnarray}}

\newcommand\comment[1]{ \hbox{[{\it Comment suppressed here.}\/]} }
\newcommand\hide[1]{}

\newcommand{\tr}{{\rm  tr}}
\newcommand{\str}{{\rm  str}}
\newcommand{\Tr}{\hbox{Tr}}
\newcommand{\Str}{\hbox{Str}}
\newcommand{\Ln}{\hbox{Ln}}
\newcommand{\inter}{{\mbox{\scriptsize int}}}
\newcommand{\Gammatpi}{\Gamma_{\mbox{\scriptsize 2PI}}}
\newcommand{\Gammaint}{\Gamma_\inter}
\newcommand{\bcG}{{\bar{\cal G}}}

\newcommand{\cG}{{\cal G}}
\newcommand{\cC}{{\cal C}}

\newcommand{\bSigma}{\bar{\Sigma}}

\newcommand{\bra}{\langle}
\newcommand{\ket}{\rangle}
\newcommand{\ketc}{\ket_{\!_c}}

\newcommand{\nn}{\nonumber \\}

\def\slashchar#1{\setbox0=\hbox{$#1$}           
   \dimen0=\wd0                                 
   \setbox1=\hbox{/} \dimen1=\wd1               
   \ifdim\dimen0>\dimen1                        
      \rlap{\hbox to \dimen0{\hfil/\hfil}}      
      #1                                        
   \else                                        
      \rlap{\hbox to \dimen1{\hfil$#1$\hfil}}   
      /                                         
   \fi}

\title{Ward Identities for the 2PI effective action in QED}

\author{Urko Reinosa\\
	Institut f{\"u}r Theoretische Physik, Universit{\"a}t Heidelberg,\\
	Philosophenweg 16, 69120 Heidelberg, Germany.\\
	E-mail: \email{u.reinosa@thphys.uni-heidelberg.de}}
\author{Julien Serreau\\
	Astro-Particule et Cosmologie,\thanks{APC is unit\'e mixte de recherche UMR 7164 (CNRS, Universit\'e Paris 7, CEA, Observatoire de Paris).} Universit\'e Paris 7 - Denis Diderot,\\
	10, rue A. Domon et L. Duquet, 75205 Paris Cedex 13, France\\
	E-mail: \email{julien.serreau@apc.univ-paris7.fr}}

\preprint{HD-THEP-07-20}	

\abstract{We study the issue of symmetries and associated Ward-like identities in the context of two-particle-irreducible (2PI) functional techniques for abelian gauge theories. In the 2PI framework, the $n$-point proper vertices of the theory can be obtained in various different ways which, although equivalent in the exact theory, differ in general at finite approximation order. We derive generalized (2PI) Ward identities for these various $n$-point functions and show that such identities are exactly satisfied at any approximation order in 2PI QED. In particular, we show that 2PI-resummed vertex functions, i.e. field-derivatives of the so-called 2PI-resummed effective action, exactly satisfy standard Ward identities. We identify another set of $n$-point functions in the 2PI framework which exactly satisfy the standard Ward identities at any approximation order. These are obtained as field-derivatives of the two-point function $\bcG^{-1}[\varphi]$, which defines the extremum of the 2PI effective action. We point out that the latter is not constrained by the underlying symmetry. As a consequence, the well-known fact that the corresponding gauge-field polarization tensor is not transverse in momentum space for generic approximations does not constitute a violation of (2PI) Ward identities.
More generally, our analysis demonstrates that approximation schemes based on 2PI functional techniques respect all the Ward identities associated with the underlying abelian gauge symmetry. Our results apply to arbitrary linearly realized global symmetries as well.
}


\begin{document} 



\section{Introduction}
\label{sec:intro}

Nonperturbative approximation schemes based on systematic expansions of the two-particle-irreducible (2PI) effective action \cite{Luttinger:1960ua,Cornwall:1974vz,DeDom:1964vz} provide powerful calculational tools in situations where standard (e.g. perturbative) expansion schemes break down. Successful applications in recent years include the thermodynamic and/or real-time properties of quantum fields in equilibrium at high temperatures \cite{Blaizot:1999ip,Berges:2004hn} (see \cite{Blaizot:2003tw} for reviews) or close to equilibrium \cite{Aarts:2003bk}, or the study of far-from-equilibrium dynamics and late-time thermalization \cite{Berges:2000ur,Berges:2002cz} (see \cite{Berges:2003pc,Berges:2004vw} for reviews).

An important issue in this context concerns the question of symmetries. One of the main concerns is that the two-point function defined as an extremum of the 2PI functional, does not satisfy standard symmetry constraints for generic approximations. For instance, in scalar theories with spontaneously broken $O(N)$ symmetry, this
two-point function does not exhibit a Goldstone mode \cite{Baym:1977qb,Ivanov:2005yj}.\footnote{See however \cite{Leupold:2006bp}.} A similar situation arises in QED, where the corresponding photon polarization tensor is not transverse in momentum space at finite approximation order, see e.g. \cite{Mottola:2003vx}.\footnote{It must be stressed however that, for systematic approximation schemes, such as for instance 2PI loop- or $1/N$-expansions, these apparent violations of standard Ward identities only occur at higher order than the approximation order, see e.g. \cite{Aarts:2002dj,vanHees:2002bv}.} 

The origin of these apparent violations of Ward identities can be traced back to the fact that, contrarily to the usual definition of the two-point vertex function as a geometrical object (i.e. the second derivative of the standard (1PI) effective action), the above-mentioned propagator is obtained from a variational procedure. 
It is important to remind that, in the 2PI framework, not only the two-point function, but all higher correlation and/or proper vertex functions of the theory can be obtained in various different ways, see for instance \cite{Berges:2005hc,Alejandro}. Of course, these different definitions are fully equivalent in the exact theory. However, this is not true in general at finite approximation order. The key point is to realize that symmetry constraints on $n$-point functions can be very different depending on which definition one is using. For instance, the two-point function can be obtained either as an extremum of the 2PI effective action, or as the second derivative of the 2PI-resummed effective action, i.e. the 2PI functional evaluated at this extremum in propagator space. These are very different objects and are thus constrained differently by the underlying symmetry.
In fact, it has been pointed out that the second derivative of the 2PI-resummed effective action does satisfy Goldstone theorem for scalar theories in the broken phase \cite{Aarts:2002dj,vanHees:2002bv} and, similarly, that the corresponding polarization tensor in abelian gauge theories is indeed transverse in momentum space \cite{Berges:2004vw}.

Another important issue in the 2PI framework, deeply related to symmetries, concerns the question of renormalization. There has been important progress in this context in recent years
\cite{vanHees:2001ik,Berges:2005hc,Reinosa:2005pj}.\footnote{See also \cite{Pawlowski:2005xe} for an approach based on
the functional renormalization group.} The basic issue is that an intrinsically nonperturbative renormalization
procedure is required in order to deal with (partial) resummations inherent to 2PI approximation schemes. The
renormalization of the variational equation for the two-point function is now well understood for theories with scalar
\cite{vanHees:2001ik,vanHees:2002bv} and/or fermionic \cite{Reinosa:2005pj} degrees of freedom and, recently, a complete
renormalization procedure for the 2PI-resummed effective action -- that is for all $n$-point proper vertex functions of the theory -- has been developed for scalar theories \cite{Berges:2005hc}. As an important result, it has been shown that global symmetries -- including spontaneously broken symmetries -- are preserved by renormalization. This generalizes the standard results of renormalization theory to 2PI renormalization.

It is important to extend these studies to theories with local symmetries. We have achieved this program for abelian
gauge theories. The present paper is the second of a series of three, where we present our results. In
\Ref{Reinosa:2006cm}, we have discussed the renormalization of the variational equations for the photon and fermion
two-point functions in QED. This requires one to introduce new counterterms, which do not appear in usual perturbation
theory, but which are allowed by the 2PI Ward identities. Therefore, the underlying gauge symmetry is preserved by
renormalization. In \Ref{QED2} we extend on this and develop a complete renormalization theory for the 2PI-resummed
effective action, generalizing the techniques put forward in \cite{Berges:2005hc} for scalar theories. As a general
result, we show that renormalization preserves the local symmetry (i.e. Ward identities) at any finite approximation
order. This yields a systematic procedure to define (UV) finite approximation schemes which respect all the symmetries
of the theory. However, before embarking on such an analysis, it is important to identify the different constrains that symmetries
of the classical theory put on the various $n$-point functions in the 2PI framework. This is the purpose of the present
paper. We restrict to the case of linearly realized symmetries and focus on the example of QED to illustrate our point.
It is crucial to distinguish the symmetry constraints on the 2PI effective action from the usual symmetry constraints on the standard (1PI) effective action. We shall generically call the former the 2PI Ward identities and the latter the 1PI or standard  Ward identities. 

In \Sec{sec:generalities}, we describe the 2PI formulation of QED, employing a superfield formalism to deal with bosonic
and fermionic fields in a unified way. We point out the importance of including mixed (e.g. boson-fermion) correlators
in the description. \Sec{sec:sym} is the core of the paper. We analyze the implications of the underlying gauge
invariance of the theory on the bare 2PI functional. This results in generalized Ward identities for the various
$n$-point proper vertex functions of the theory. We focus on two particular families of $n$-point functions in the 2PI
framework: we call the first one the set of ``2PI-resummed vertex functions'', defined as field derivatives of the 2PI-resummed effective action, and the second one the set of ``2PI vertex functions'', obtained as field-derivatives of the inverse propagator $\bcG^{-1}[\varphi]$, which defines the extremum of the 2PI effective action in propagator space. Equivalently, these 2PI vertex functions can be obtained as solutions of appropriate Bethe-Salpeter-like equations.
 
We show that the former, the 2PI-resummed vertex functions, exactly satisfy standard Ward identities, as is to be expected from their geometrical interpretation in field space. This generalizes the standard result for (1PI) vertex functions. As a more unexpected result, we find that, except for the two-point function $\bcG[\varphi]$ itself, the 2PI vertex functions also exactly satisfy the standard Ward identities. This is one of the central results of this paper. It plays a crucial role in the renormalization program \cite{QED2}. It is also of particular importance for various possible applications of 2PI methods to abelian gauge theories. For instance, this yields an efficient way to construct systematic non-perturbative approximations to the QED photon-fermion-antifermion vertex which satisfies the usual Ward identity. This guarantees for instance, that standard low-energy theorems \cite{BjorkenDrell} are satisfied at any approximation order. It is also important in the context of transport coefficients calculations, such as the QED electrical conductivity, see e.g.~\cite{Aarts:2002tn}. Finally, this provides a useful guide to devise gauge-invariant truncations of Schwinger-Dyson equations.

Our analysis also reveals that 2PI Ward identities do not put any direct constraint on the two-point function
$\bcG[\varphi]$, but only on its field derivatives. For instance, as already noticed earlier \cite{Reinosa:2006cm}, the photon polarization tensor obtained from the extremum of the 2PI effective action in QED is not constrained to be transverse in momentum space at any finite approximation order. Conversely, the nontransversality of the latter should not be interpreted as a violation of (2PI) Ward identities. In other words, the 2PI functional has no reason to satisfy the 1PI Ward identities. The reason for the ``variational'' polarization tensor to become transverse in the exact theory is that it gets identical to the ``geometrical'' polarization tensor, i.e. the one obtained from the 2PI-resummed two-point function, which is indeed constrained to be transverse by Ward identities.

For the above results to be useful at all, they should hold for practical approximations. In \Sec{sec:gaugeinvapprox},
we discuss the symmetry properties of the standard diagrammatic expansion of the 2PI effective action
\cite{Cornwall:1974vz}. We show how to systematically construct gauge-invariant approximation schemes -- i.e. which
preserve the (2PI) Ward identities to all orders of approximation. For the particular case of QED, we find that any approximation is gauge-invariant in this sense. 

A more general discussion of nonlinear (local) symmetries and corresponding Slavnov-Taylor identities for $n$PI effective actions, as well as some technical material and a discussion of particular aspects of scalar QED are given in the appendices. Finally, we mention that, although the analysis presented in this paper concerns the case of abelian gauge theories, our results hold for arbitrary theories with linear global symmetries, including spontaneously broken symmetries.

\section{The 2PI effective action in QED}
\label{sec:generalities}

\subsection{Generalities and superfield formalism}

We consider QED in the covariant gauge and use dimensional regularization. The gauge-fixed classical action reads, with standard notations,
\begin{equation}
\label{eq:classact}
S[A,\psi,\bar\psi]=\int_x \left\{\bar\psi\Big[i\slashchar{\partial}-e\slashchar{A}-m\Big]\psi+\frac{1}{2}A^\mu\Big[g_{\mu\nu}\partial^2-(1-\lambda)\partial_\mu\partial_{\nu}\Big]A^\nu\right\}\,,
\end{equation}
where $\int_x\equiv\int d^dx$ and $\lambda$ is the gauge-fixing parameter. Aside from the gauge-fixing term, the classical action is invariant under the gauge transformation
\begin{equation}\label{eq:gauge}
\psi(x)\rightarrow e^{i\alpha(x)}\psi(x)\,,\;\;
\bar\psi(x)\rightarrow e^{-i\alpha(x)}\bar\psi(x)\,,\;\;
A_\mu(x)\rightarrow A_\mu(x)-\frac{1}{e}\,\partial_\mu\alpha(x)\,,\nonumber\\
\end{equation}
where $\alpha(x)$ is an arbitrary real function. 
The free inverse fermion and photon propagators are given by
\begin{eqnarray}
\label{eq:freepf}
iD_{0,\bar\alpha\alpha}^{-1}(x,y) & = & \left[\,i\slashchar{\partial}_x-m\,\right]_{\bar\alpha\alpha}\delta^{(4)}(x-y)\,,\\
\label{eq:freepb}
iG_{0,\mu\nu}^{-1}(x,y) & = & \left[\,g_{\mu\nu}\partial_x^2-
(1-\lambda)\partial^x_\mu\partial^x_\nu\,\right]\delta^{(4)}(x-y)\,.
\end{eqnarray}

It is convenient to grab the bosonic and fermionic fields $A$, $\psi$ and $\bar\psi$ in a $12$-component superfield
\bea
\label{eq:sfield}
 \varphi\equiv\left(
 \begin{tabular}{c}
  $A$\\
  $\psi$\\
  $\bar\psi^t$
  \end{tabular}
  \right),
\eea
where $t$ stands for transposition, here of Dirac indices. We assign a fermion number $q_m$ to each component $\varphi_m$ of $\varphi$ such that $\smash{q_m=0}$ for bosonic ($A$-like) components, $\smash{q_m=+1}$ for fermionic $\psi$-like components and $\smash{q_m=-1}$ for fermionic $\bar\psi$-like components. Also, for simplicity, we employ a notation where the space-time variables, if not written explicitly, are put together with the Lorentz, or Dirac, or superfield indices.
Repeated indices are implicitly summed over, which includes an integration over space-time variables. 

Since we deal with Grassmann variables, we may have to distinguish between left (L) and right (R) derivatives, defined as follows: The variation $\delta F$ of a given functional $F[\varphi]$ under an arbitrary variation $\delta\varphi$ of its argument is given by
\beq
 \delta F=\frac{\delta_L F}{\delta\varphi_n}\,\delta\varphi_n=\delta\varphi_n\,\frac{\delta_R F}{\delta\varphi_n}\,.
\eeq
In the present paper, we mostly use right derivatives. Therefore, we adopt the convention that, unless explicitly specified, functional derivatives are understood as right derivatives. Moreover, successive (right) derivatives are noted such as the rightmost derivative acts first:
\beq
 F[\varphi+\delta\varphi]=\sum_{n\ge 0}\frac{1}{n!}\,\delta\varphi_1\ldots\delta\varphi_n\,\frac{\delta^{n}F}{\delta\varphi_n\ldots\delta\varphi_1}\,.
\eeq

We write the classical action \Eqn{eq:classact} as
\beq
\label{eq:saction}
 S[\varphi]=S_0[\varphi]+S_{\rm int}[\varphi]
\eeq
with free (quadratic) part
\beq
\label{eq:freeact}
 S_0[\varphi]=\frac{1}{2}\,\varphi_m i\cG_{0,mn}^{-1}\,\varphi_n\,,
\eeq
and interaction part $S_{\rm int}[\varphi]$. \Eqn{eq:freeact} defines the inverse free propagator $\cG_0^{-1}$. One has equivalently
\beq
\label{eq:sfreeprop}
 i\cG_{0,mn}^{-1}\equiv\left.(-1)^{q_n}\frac{\delta^{2} S[\varphi]}{\delta\varphi_m\delta\varphi_n}\right|_{\varphi=0}\,.
\eeq
The only nonvanishing components of $\cG_0^{-1}$ are given by Eqs.~\eqn{eq:freepf}-\eqn{eq:freepb}:
\bea
\label{eq:scor0}
 i\cG_0^{-1}=\left(
 \begin{tabular}{ccc}
 $iG_0^{-1}$&$0$&$0$\\
 $0$&$0$&$\left(-iD_0^{-1}\right)^t$\\
 $0$&$iD_0^{-1}$&$0$
 \end{tabular}
 \right),
\eea
where the symbol $t$ for transposition includes space-time variables.

To define the 2PI effective action, one introduces the following generating functional \cite{Cornwall:1974vz}:
\beq
\label{eq:genfunc}
 {\rm e}^{iW[{\cal J},\,{\cal K}]}=\int{\cal D}\hat\varphi\,{\rm e}^{i\{S[\hat\varphi]+\hat\varphi_m\,{\cal J}_m+\frac{1}{2}\hat\varphi_m\hat\varphi_n\,{\cal K}_{nm}\}}\,,
\eeq
with linear and bilinear external sources ${\cal J}$ and ${\cal K}$. Notice that the components ${\cal J}_m$ and ${\cal K}_{mn}$ of these classical sources are either usual (commuting), or Grassmann (anti-commuting) functions, as appropriate. It follows from \Eqn{eq:genfunc} that the bilinear source ${\cal K}$ has the following property under transposition in superfield space:
\beq
\label{eq:transposource}
 {\cal K}_{mn}=(-1)^{q_{m}q_{n}}{\cal K}_{nm}\,.
\eeq
The connected one- and two-point functions $\varphi$ and $\cG$ in the presence of sources are defined as
\beq
\label{eq:onepointdef}
 \frac{\delta_L W}{\delta {\cal J}_m}\equiv\varphi_m
\eeq
and
\beq
\label{eq:twopointdef}
 \frac{\delta_L W}{\delta {\cal K}_{nm}}\equiv \frac{1}{2}\,\left(\varphi_m\,\varphi_n+\cG_{mn}\right)\,.
\eeq
The 2PI effective action is the double Legendre transform of the generating functional $W[{\cal J},{\cal K}]$:
\beq
\label{eq:legendre}
 \Gamma_{\rm 2PI}[\varphi,\cG]=W[{\cal J},{\cal K}]-\frac{\delta_L W}{\delta {\cal J}_m}\,{\cal J}_m-\frac{\delta_L W}{\delta {\cal K}_{mn}}\,{\cal K}_{mn}\,.
\eeq
One has the obvious relations:
\beq
\label{eq:source1}
 \frac{\delta_R \Gamma_{\rm 2PI}}{\delta \varphi_{m}}=-{\cal J}_m-\varphi_n{\cal K}_{nm}
\eeq
and
\beq
\label{eq:source2}
 \frac{\delta_R \Gamma_{\rm 2PI}}{\delta \cG_{mn}}=-\frac{1}{2}\,{\cal K}_{nm}\,.
\eeq

Using the decomposition \eqn{eq:saction} of the classical action, the 2PI functional can be pa\-ra\-met\-ri\-zed as (see \cite{Luttinger:1960ua,Cornwall:1974vz}  and App.~\ref{appsec:gaussian})
\beq 
\label{eq:2PI} \Gammatpi[\varphi,\cG]=S_0[\varphi]+\frac{i}{2}\,\Str\,\Ln\cG^{-1}+\frac{i}{2}\,\Str\,\cG_0^{-1}\cG+\Gammaint[\varphi,\cG]\,,
\eeq
where $\Str$ denotes the functional supertrace (see App.~\ref{appsec:gaussian}), $\cG_0^{-1}$ is the free inverse
propagator defined in \Eqn{eq:sfreeprop} and $\Gamma_{\rm int}[\varphi,\cG]$ is the set of closed
two-particle-irreducible (2PI) diagrams -- up to an overall factor $(-i)$ -- with lines corresponding to $\cG$ and vertices obtained from the shifted action $S_{\rm int}[\varphi +\hat\varphi]$, where $\hat\varphi$ is the integration variable in the path integral representation \eqn{eq:genfunc}.

The physical correlator $\bcG[\varphi]$ in the presence of a nonvanishing field $\varphi$ is obtained for vanishing bilinear sources $\smash{{\cal K}=0}$, which corresponds to the stationarity condition, see \Eqn{eq:source2},
\beq
\label{eq:stat}
 \left.\frac{\delta\Gammatpi[\varphi,\cG]}{\delta \cG}\right|_{\bcG[\varphi]}=0\,.
\eeq
Using the properties of the supertrace (see App.~\ref{appsec:gaussian}), this can be written as
\beq
\label{eq:EOM}
 \bcG^{-1}[\varphi]=\cG^{-1}_0-\bar\Sigma[\varphi]\,,
\eeq
where the components of the self-energy $\bar\Sigma[\varphi]$ are given by
\beq
\label{eq:sigmadef}
 \bar\Sigma_{mn}[\varphi]\equiv \left.(-1)^{q_n}\,2i\,\frac{\delta\Gammaint[\varphi,\cG]}{\delta \cG_{nm}}\right|_{\bcG[\varphi]}\,.
\eeq

Finally, the effective action $\Gamma[\varphi]$, the generating functional for 1PI $n$-point vertex functions, is obtained as 
\beq
\label{eq:1PI}
 \Gamma[\varphi]\equiv\Gammatpi[\varphi,\bcG[\varphi]]\,.
\eeq
This defines the 2PI-resummed effective action.
The above equation is a trivial identity in the exact theory. For finite approximations however, Eqs.~\eqn{eq:2PI}, \eqn{eq:stat} and \eqn{eq:1PI} define an efficient way to devise systematic nonperturbative approximations of the effective action $\Gamma[\varphi]$ through systematic expansions of the functional $\Gamma_{\rm int}[\varphi,\cG]$ in \Eqn{eq:2PI} \cite{Berges:2005hc}.

\subsection{A note on mixed two-point correlators}

As described above, in order to define the 2PI effective action in full generality, one needs to introduce all possible
bilinear sources in \eqn{eq:genfunc} or, equivalently, all two-point correlators for Maxwell and Dirac fields (see also \cite{Cooper:2002qd,Arrizabalaga:2002hn,Calzetta:2004sh}). These include the usual ones:
\beq
\label{eq:cor1}
 G_{\mu\nu}(x,y)\equiv\Big< A_\mu(x) A_\nu(y)\Big>_{\!c}\quad;\quad D_{\alpha\bar\beta}(x,y)\equiv\Big<
 \psi_{\alpha}(x)\bar\psi_{\bar\beta}(y)\Big>_{\!c}\,,
\eeq
as well as the mixed correlators:
\beq
\label{eq:cor2}
 K_{\alpha\nu}(x,y)\equiv\Big< \psi_\alpha(x)A_\nu(y)\Big>_{\!c}\quad;\quad\bar K_{\mu\bar\beta}(x,y)\equiv\Big< A_\mu(x)\bar\psi_{\bar\beta}(y)\Big>_{\!c}
\eeq
and
\beq
\label{eq:cor3}
 F_{\alpha\beta}(x,y)\equiv\Big< \psi_\alpha(x)\psi_\beta(y)\Big>_{\!c}\quad;\quad\bar F_{\bar\alpha\bar\beta}(x,y)\equiv\Big< \bar\psi_{\bar\alpha}(x)\bar\psi_{\bar\beta}(y)\Big>_{\!c}\,.
\eeq
Here, the brackets denote 
expectation values in presence of external sources:\footnote{In terms of field operators, the above correlation functions should involve a bosonic/fermionic $T$-product. This, however, is not needed in the path integral formulation employed here.}
\beq
\label{eq:expvalue}
 \Big< F[\hat\varphi]\Big>\equiv\frac{\int{\cal D}\hat\varphi\,F[\hat\varphi]\,{\rm
 e}^{i\{S[\hat\varphi]+\hat\varphi_m\,{\cal J}_m+\frac{1}{2}\hat\varphi_m\hat\varphi_n\,{\cal K}_{nm}\}}}{\int{\cal
 D}\hat\varphi\,{\rm e}^{i\{S[\hat\varphi]+\hat\varphi_m\,{\cal J}_m+\frac{1}{2}\hat\varphi_m\hat\varphi_n\,{\cal K}_{nm}\}}}\,,
\eeq
and the subscript $c$ stands for connected expectation values.
Notice that $G$, $D$, $F$ and $\bar F$ are usual commuting functions, whereas $K$ and $\bar K$ are anticommuting (Grassmann) functions. 

The various correlators \eqn{eq:cor1}-\eqn{eq:cor3} can be put together as components of the connected correlator of
superfields
\beq
\label{eq:sprop}
 \cG\equiv\Big< \hat\varphi\,\hat\varphi^t\Big>_{\!c}\,,
\eeq 
or, in components, $\smash{\cG_{mn}=\bra \hat\varphi_m\hat\varphi_n\ketc}$. One has explicitly,
\bea
\label{eq:scor}
 \cG=\left(
 \begin{array}{ccc}
 G&K^t&\bar K\\
 K&F&D\\
 \bar K^t&-D^t&\bar F
 \end{array}
 \right)\,.
\eea
Notice that under transposition, see also \Eqn{eq:transposource},
\beq
\label{eq:symprop}
 \cG_{mn}=(-1)^{q_mq_n}\cG_{nm}\,.
\eeq 
It follows that the inverse propagator has the symmetry property
\beq
\label{eq:syminv}
 \cG^{-1}_{mn}=(-1)^{q_mq_n+q_m+q_n}\cG^{-1}_{nm}\,.
\eeq
One easily checks, using Eqs.~(\ref{eq:sfreeprop}) and (\ref{eq:sigmadef}), that both the free inverse propagator $\cG_0^{-1}$ and the self-energy $\bar\Sigma[\varphi]$ behave as inverse propagators under transposition, as they should, see \Eqn{eq:EOM}. One has, explicitly,
\bea
\label{eq:ssigm}
 \bar\Sigma=\left(
 \begin{array}{ccc}
 \bar\Sigma_{AA}&\bar\Sigma_{A\psi}&-\bar\Sigma_{\bar\psi A}^t \\
 -\bar\Sigma_{A\psi}^t& \bar\Sigma_{\psi\psi}& -\bar\Sigma_{\bar\psi\psi}^t\\
 \bar\Sigma_{\bar\psi A}&\bar\Sigma_{\bar\psi\psi}& \bar\Sigma_{\bar\psi\bar\psi}
 \end{array}
 \right)\,.\nonumber
\eea

All mixed correlators vanish in the absence of external sources\footnote{\label{ft:u1} More precisely, in the absence of fermionic and mixed sources, i.e. ${\cal J}_m=0$ for $q_m\neq0$ and ${\cal K}_{mn}=0$ for $q_m+q_n\neq0$, the theory has a global $U(1)$ symmetry, which implies, in particular, that $\varphi_m=0$ for $q_m\neq0$ and $\cG_{mn}=0$ for $q_m+q_n\neq0$.}. But it is important to stress that their role is crucial in intermediate calculations involving functional derivatives. For instance the function $\delta\Gammatpi/\delta K\delta\bar K$ does not vanish for ${\cal K}=0$. Only in a few specific cases, when one restricts one's attention to simple particular  vertex functions, do these mixed correlators play no role and can be discarded from the beginning, that is directly at the level of the 2PI effective action (see below).

\subsection{2PI and 2PI-resummed vertex functions}

In the 2PI framework, $n$-point vertex functions can be obtained in different ways, see e.g.
\cite{Aarts:2002dj,vanHees:2002bv,Berges:2005hc,Alejandro}, all of which are of course strictly equivalent in the exact
theory, but which differ in general for finite approximations. The most straightforward definition is via the usual $n$-th derivatives of the (2PI-resummed) effective action \eqn{eq:1PI}:
\beq
\label{eq:npoint1}
 \Gamma^{(n)}_{1\ldots n}\equiv\left.\frac{\delta^n\Gamma[\varphi]}{\delta\varphi_n\cdots\delta\varphi_1}\right|_{\bar\varphi}\,,
\eeq
taken at $\smash{\varphi=\bar\varphi}$, the physical value of the field, obtained from the stationarity condition
\beq
\label{eq:barphi}
 \left.\frac{\delta\Gamma[\varphi]}{\delta\varphi_1}\right|_{\bar\varphi}=0\,,
\eeq
which corresponds to vanishing external sources, see Eqs.~(\ref{eq:source1}) and (\ref{eq:source2}).\footnote{For (gauge fixed) QED in $C$-invariant physical states, $\bar\varphi=0$.}
We shall refer to the vertex functions \eqn{eq:npoint1}, obtained from the 2PI-resummed effective action \eqn{eq:1PI}, as the 2PI-resummed vertex functions. 

Other possible definitions of $n$-point vertex functions involve derivatives of the 2PI generating functional \eqn{eq:2PI} with respect to the two-point function $\cG$, see e.g. \cite{McKay:1989rk,Aarts:2003bk,vanHees:2002bv,Berges:2005hc,Alejandro}. For instance, the two-point vertex function, which is nothing but the field self-energy can either be obtained from the second derivative of the 2PI-resummed effective action, as in \Eqn{eq:npoint1} above, or directly from \Eqn{eq:sigmadef}. In turn, higher $n$-point functions can be obtained as derivatives of the self-energy $\bar\Sigma[\varphi]$:
\beq
\label{eq:npoint2}
 iV^{(p+2)}_{mn;1\cdots p}\equiv\left.(-1)^{q_m}\frac{\delta^p\bar\Sigma_{nm}[\varphi]}{\delta\varphi_{p}\cdots\delta\varphi_{1}}\right|_{\bar\varphi}\,.
\eeq
The fact that $V^{(p+2)}$ defines the $(p+2)$-point vertex function directly follows from the relation:
\beq
\label{eq:identity}
 (-1)^{q_n}\frac{\delta^{2}\Gamma[\varphi]}{\delta\varphi_m\delta\varphi_n}=i\bcG^{-1}_{mn}[\varphi]\,,
\eeq
which holds for arbitrary $\varphi$ in the exact theory, as shown in App.~\ref{sec:appvertex} (see also
\cite{Cornwall:1974vz}). In the exact theory, one has therefore
\beq
\label{eq:vertexidentity}
 V^{(p+2)}_{mn;1\cdots p}=\Gamma^{(p+2)}_{mn1\ldots p}\,,
\eeq
for $\smash{p\ge0}$. We shall refer to the functions $V^{(n)}$, defined in \Eqn{eq:npoint2} for $\smash{n\ge2}$, as the 2PI $n$-point vertex functions.\footnote{We stress that the name ``2PI vertex function'' is meant to recall the definition of these functions. It should be kept in mind that these are really proper vertex functions and are, therefore, one-particle-irreducible, not two-particle-irreducible objects.} There are other possible definitions of higher order vertex functions involving more than one derivative of the 2PI effective action with respect to the two-point function $\cG$ \cite{McKay:1989rk,Aarts:2003bk,Berges:2005hc,Alejandro}. We shall not consider these in the present paper.

At finite approximation order, the identity \eqn{eq:identity} -- and therefore \eqn{eq:vertexidentity} -- is not satisfied in general and thus the various definitions of $n$-point vertex functions, e.g. Eqs.~\eqn{eq:npoint1} and \eqn{eq:npoint2}, differ. Notice however that for systematic 2PI approximation schemes, such as for instance a loop- or a $1/N$-expansion, the identity \eqn{eq:identity} is only violated at higher order than the approximation order, see e.g. \cite{Aarts:2002dj,vanHees:2002bv,Berges:2005hc,Alejandro}. Therefore, the different definitions of $n$-point vertex functions all agree with each other at the approximation order.\footnote{This is for instance crucial for the renormalization program \cite{Berges:2005hc}.} 

The 2PI-resummed vertex functions \eqn{eq:npoint1} can be directly related to the 2PI vertex functions \eqn{eq:npoint2} through \Eqn{eq:1PI}. Indeed, derivatives of the 2PI-resummed effective action $\Gamma[\varphi]$ implicitly involve derivatives of $\bcG[\varphi]$,
which can be directly related to the 2PI vertex functions \eqn{eq:npoint2} through \Eqn{eq:EOM}. 
Here, we illustrate this point focusing on cubic theories -- relevant for the present QED case -- for which various simplifications occur (see \cite{QED2} for a detailed discussion).
For a cubic interaction, we write the interaction term in the classical action as 
\beq
\label{eq:sclassint}
 S_{\rm int}[\varphi]=\frac{1}{3!}\,\lambda_{mnp}\,\varphi_m\varphi_n\varphi_p\,,
\eeq
where the classical vertex $\lambda$ is such that, for each permutation of a pair of neighboring indices, $\smash{\lambda_{mnp}=(-1)^{q_mq_n}\lambda_{nmp}}$, etc. Notice also that Lorentz invariance implies that $\smash{\lambda_{mnp}\neq 0}$ only if $\smash{q_m+q_n+q_p=0}$. Thus the only nonvanishing components of $\lambda$ are $c$-numbers. For such theories, it is easy to see that the $\varphi$-dependence of $\Gammaint[\varphi,\cG]$ is all contained in the zero-loop (classical) and one-loop contributions:\footnote{For cubic theories, the only closed 2PI diagrams one can construct with the $\varphi$-dependent vertices of the shifted action $S[\varphi+\hat\varphi]$ have either zero (classical vertex) or one loop, see also \Sec{sec:gaugeinvapprox} below.} 
\beq
\label{eq:sfielddep}
 \Gammaint[\varphi,\cG]=S_{\rm int}[\varphi]+\Gammaint^{\rm 1-loop}[\varphi,\cG]+ \Gamma_2[\cG]\,,
\eeq
where
\beq
\label{eq:s1loop}
 \Gammaint^{\rm 1-loop}[\varphi,\cG]=\frac{1}{2}\,\lambda_{mnp}\,\varphi_m\,\cG_{np}\,.
\eeq
Here, $\Gamma_2[\cG]$ is the sum of $n$-loop closed 2PI diagrams with $\smash{n\ge 2}$, with lines given by $\cG$ and vertices given by \Eqn{eq:sclassint}. Therefore, the first derivative of the 2PI-resummed effective action has the simple exact expression
\beq
\label{eq:firstder}
\frac{\delta\Gamma}{\delta\varphi_1}=\left.\frac{\delta\Gammatpi}{\delta\varphi_1}\right|_\bcG=\frac{\delta S}{\delta\varphi_1}+\frac{1}{2}\lambda_{1mn}\bcG_{mn}[\varphi]\,,
\eeq
where we used the stationarity condition \eqn{eq:stat} in writing the first equality.
Differentiating once more with respect to the field, one obtains the following expression for the 2PI-resummed two-point function (see \cite{QED2} for details), 
\beq
\label{eq:twopointtt} \left.\frac{\delta^{2}\Gamma}{\delta\varphi_2\delta\varphi_1}\right|_{\bar\varphi}=(-1)^{q_1}i\cG_{0,21}^{-1}+\frac{1}{2}\lambda_{1mn}\left(\bcG\,\frac{\delta\bSigma}{\delta\varphi_2}\,\bcG\right)_{\!\!\!mn},
\eeq
where the term between brackets is to be interpreted as a matrix product and it is also understood that the RHS is to be evaluated for $\smash{\varphi=\bar\varphi}$. Here, we explicitly used the fact that $\smash{\bar\varphi=0}$.\footnote{We also used the fact that in the absence of external fermionic and mixed sources, the theory has a global $U(1)$ symmetry corresponding to net fermion number conservation. As a consequence, any quantity having a nonvanishing fermion number vanishes. For instance, the mixed components of the correlator vanish; also $\delta\bSigma_{mn}/\delta\varphi_p=0$ unless $q_m+q_n+q_p=0$, etc.}

Taking further derivatives with respect to the field $\varphi$, one easily obtains explicit expressions for higher 2PI-resummed vertex functions in terms of 2PI-vertices \eqn{eq:npoint2}. As clear from the above example, 2PI-resummed vertices of a given order involve higher-order 2PI vertices in general. For instance, to compute the 2PI-resummed two-point function $\Gamma^{(2)}$ through \Eqn{eq:twopointtt}, one needs the 2PI three-point vertex $V^{(3)}\propto\delta\bar\Sigma/\delta\varphi$. As for 2PI vertices themselves,  they are solution of integral, Bethe-Salpeter-like equations \cite{McKay:1989rk,Aarts:2003bk,Berges:2005hc,Alejandro,QED2}. For instance the 2PI three-point vertex discussed here satisfies the following equation \cite{QED2}:
\beq
\label{eq:integral}
 \left.(-1)^{q_n}\frac{\delta\bSigma_{mn}}{\delta\varphi_p}\right|_{\bar\varphi}=i\lambda_{nmp}+\left(\bcG\,\frac{\delta\bar\Sigma}{\delta\varphi_p}\,\bcG\right)_{\!\!\!ab}\left.\frac{2i\delta\Gammaint}{\delta\cG_{ab}\delta\cG_{nm}}\right|_{\bcG},
\eeq
where it is again implicitly understood that $\varphi=\bar\varphi$ on the RHS of the equality. Similar equations can be
derived for higher-order 2PI vertices. It is to be noticed that the equation for a given 2PI vertex only involves 2PI vertices of equal or lower order (see e.g. \cite{Berges:2005hc,QED2,Alejandro}).

\begin{figure}[tbp]
\begin{center}
\includegraphics[width=10.cm]{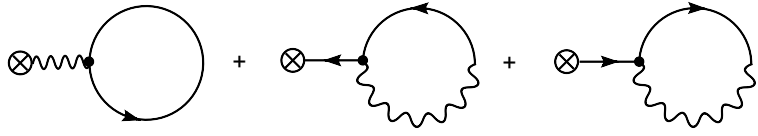}
\caption{\label{fig:1loopQED} Explicit one-loop contributions to $i\Gammaint[\varphi,\cG]$ in QED, see
Eq.~(2.45). The crosses
represent the fields $A$, $\psi$ or $\bar\psi$; The straight line in the first graph represents the fermion correlator
$D\equiv\langle \psi\bar\psi\rangle$, while the mixed wavy-straight lines in the last two diagrams represent the mixed
boson-fermion correlators $K\equiv\langle \psi A\rangle$ and $\bar K\equiv\langle A\bar\psi\rangle$. Arrows are
associated with a $\psi$ if they point towards a vertex and with a $\bar\psi$ otherwise. The black dot represents the
usual QED vertex. Together with the contribution from the classical action, this gives the exact field-dependence of the
2PI effective action in QED.}
\end{center}
\end{figure}

\begin{figure}[tbp]
\begin{center}
\includegraphics[width=12.5cm]{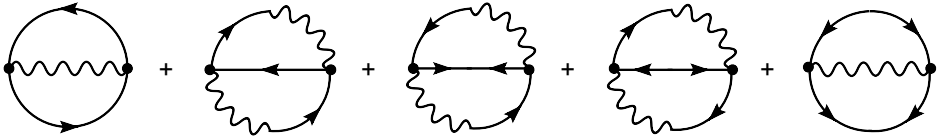}
\caption{\label{fig:2loopQED} Leading (two-loop) contributions to $i\Gamma_2[\cG]$ in QED in a loop expansion. Wavy
lines represent the gauge field propagator $G\equiv\langle AA\rangle$; Straight lines represent the fermion correlators
$D\equiv\langle \psi\bar\psi\rangle$, $F\equiv\langle \psi\psi\rangle$ and $\bar F\equiv\langle
\bar\psi\bar\psi\rangle$; Finally, mixed wavy-straight lines represent the mixed boson-fermion correlators
$K\equiv\langle \psi A\rangle$ and $\bar K\equiv\langle A\bar\psi\rangle$. The convention for arrows is the same as in
Fig.~1
}
\end{center}
\end{figure}

It is instructive to write the above equations in terms of the components of the superfields in the case of QED. The only nonvanishing components of the classical vertex are the various permutations of 
\beq
\label{eq:classvertexQED}
 \lambda_{A_\mu\bar\psi_{\bar\alpha}\psi_\alpha}(x,y,z)=-e\gamma^\mu_{\bar\alpha\alpha}\,\delta^{(4)}(x-y)\,\delta^{(4)}(x-z)\,,
\eeq 
where we made explicit  the space-time dependence. The interaction part of the 2PI effective action is given by \Eqn{eq:sfielddep} with the classical term
\beq
\label{eq:class}
 S_{\rm int}[A,\psi,\bar\psi]=-e\int_x\bar\psi(x)\slashchar{A}(x)\psi(x)\,,
\eeq
and the one-loop term
\beq
\label{eq:1loop}
 \Gammaint^{\rm 1-loop}[A,\psi,\bar\psi,{\cal G}]=e\int_x \Big\{\tr [\slashchar{A}(x) D(x,x)]-\bar\psi(x)\,\slashchar{K}(x,x)-\slashchar{\bar K}(x,x)\psi(x)\Big\}\,.
\eeq
This is represented diagrammatically in Fig.~\ref{fig:1loopQED}. The functional $\Gamma_2[\cG]$ is the sum of closed 2PI
diagrams with the usual QED vertex and with lines given by the various components of $\cG$, see \Eqn{eq:scor}. The first
(two-loop) contributions are represented in Fig.~\ref{fig:2loopQED}. 

To distinguish photon and fermion legs, it is useful to employ the following notation for the 2PI-resummed vertex function with $2m$ fermion and $n$ photon legs:\footnote{We employed a slightly different notation in \Ref{Reinosa:2006cm}.}
\beq
\label{eq:defvertex}
 \Gamma^{(2m,n)}_{\bar\alpha_1\cdots\bar\alpha_m,\alpha_1\cdots \alpha_m,\mu_1\cdots \mu_n}\equiv\Gamma^{(2m+n)}_{\bar\psi_{\bar \alpha_1}\cdots\bar\psi_{\bar\alpha_m}\psi_{\alpha_1}\cdots\psi_{\alpha_m}\bar A^{\mu_1}\cdots A^{\mu_n}}\,.
\eeq

For instance, the photon and fermion two-point functions are written as $\Gamma_{\mu\nu}^{(0,2)}\equiv\Gamma^{(2)}_{A^\mu A^\nu}=\delta\Gamma[\varphi]/\delta A^\nu\delta A^\mu|_{\bar\varphi=0}$ and $\Gamma_{\bar\alpha\alpha}^{(2,0)}\equiv\Gamma^{(2)}_{\bar\psi_{\bar\alpha}\psi_\alpha}=\delta\Gamma[\varphi]/\delta \psi_\alpha\delta \bar\psi_{\bar\alpha}|_{\bar\varphi=0}$ respectively. The latter are obtained from \Eqn{eq:twopointtt} as, see \Ref{QED2},
\beq
\label{eq:twoptb}
 \Gamma_{\mu\nu}^{(0,2)}(x,y)=iG_{0,\mu\nu}^{-1}(x,y)+e\int_{uv}\tr\left(\gamma_{\mu}\, \bar D(x,u)\, \frac{\delta\bar\Sigma_{\bar\psi\psi}(u,v)}{\delta A^\nu(y)} \,\bar D(v,x)\right),
\eeq
where the trace is over Dirac indices, and
\beq
\label{eq:twoptf}
 \Gamma_{\bar\alpha\alpha}^{(2,0)}(x,y)=iD_{0,\bar\alpha\alpha}^{-1}(x,y)-e\int_{uv}\gamma^\mu_{\bar\alpha\beta}\left(\bar D(x,u)\, \frac{\delta\bar\Sigma_{\bar\psi A}(u,v)}{\delta \psi_\alpha(y)} \,\bar G(v,x)\right)_{\!\!\!\beta\mu}\!.
\eeq
The 2PI photon-fermion vertex functions $iV^{(3)}_{\psi\bar\psi;A}=\delta\bar\Sigma_{\bar\psi\psi}/\delta A$ and $iV^{(3)}_{A\bar\psi;\psi}=\delta\bar\Sigma_{\bar\psi A}/\delta\psi$ appearing on the RHS of Eqs.~\eqn{eq:twoptb} and \eqn{eq:twoptf} respectively satisfy the following integral equations \cite{QED2}, see \Eqn{eq:integral}:
\beq
\label{eq:integralcomp1}
 \frac{\delta\bar\Sigma_{\bar\psi\psi}^{\bar\alpha\alpha}}{\delta A^\mu}=-ie\gamma^\mu_{\bar\alpha\alpha}+\left(\bar D\,\frac{\delta\bar\Sigma_{\bar\psi\psi}}{\delta A^\mu}\,\bar D\right)_{\!\!\!\beta\bar\beta}\left.\frac{-i\delta^{2}\Gammaint}{\delta D_{\beta\bar\beta}\delta D_{\alpha\bar\alpha}}\right|_\bcG,
\eeq
and
\beq
\label{eq:integralcomp2}
 \frac{\delta\bar\Sigma_{\bar\psi A}^{\bar\alpha\mu}}{\delta \psi_\alpha}=-ie\gamma^\mu_{\bar\alpha\alpha}+\left(\bar D\,\frac{\delta\bar\Sigma_{\bar\psi A}}{\delta\psi_\alpha}\,\bar G\right)_{\!\!\!\beta\nu}\left.\frac{i\delta^{2}\Gammaint}{\delta K_{\beta\nu}\delta\bar K_{\mu\bar\alpha}}\right|_\bcG,
\eeq
where we leave space-time variables implicit for simplicity\footnote{It is understood that the first terms on the RHS of
both Eqs.~(\ref{eq:integralcomp1}) and (\ref{eq:integralcomp2}), i.e. the classical vertex, is local in space-time. We do not
write explicitly the corresponding delta functions (see \Eqn{eq:classvertexQED}) for notational convenience.} In obtaining Eqs.~\eqn{eq:integralcomp1}-\eqn{eq:integralcomp2}, we have assumed $C$-invariance.
It is implicitly understood that the various terms in Eqs.~\eqn{eq:twoptb}-\eqn{eq:integralcomp2} are to be evaluated for vanishing fields $A=0$, $\psi=0$ and $\bar\psi=0$.

It is remarkable that Eqs.~\eqn{eq:twoptb} and \eqn{eq:twoptf} have a very similar structure as standard Schwinger-Dyson
equations for QED two-point functions. Here, in place of 2PI-resummed two- and three-point vertex functions which would
appear on the RHS for Schwinger-Dyson equations, appear the corresponding 2PI vertex functions, defined
previously.\footnote{This is also true for higher 2PI-resummed vertex functions. This is rooted in the fact that the
interaction term in the classical action is purely cubic.} This would make no difference in the exact theory, where
2PI-resummed and 2PI vertex functions are identical. However it does make a difference with standard Schwinger-Dyson
equations at finite approximation order. For instance, a remarkable feature of the present equations is that the
three-point vertex functions appearing on the RHS's satisfy closed integral equations,
Eqs.~\eqn{eq:integralcomp1}-\eqn{eq:integralcomp2}, and do not involve higher order vertex functions, as would be the case in standard Schwinger-Dyson hierarchy. 

Notice that \Eqn{eq:twoptf} for the fermion two-point function $\Gamma^{(2,0)}$ illustrates the importance of keeping mixed correlators in intermediate calculations. Indeed, although $\bar\Sigma_{\bar\psi A}$ vanishes in the absence of sources, the 2PI three-point function $\delta\bar\Sigma_{\bar\psi A}/\delta \psi$ is nonzero. Not including mixed correlators in the 2PI generating functional would lead to a trivial (free) expression for $\Gamma^{(2,0)}$, \Eqn{eq:twoptf}.

\section{Symmetries}
\label{sec:sym}

\subsection{2PI Ward identities}

Apart from the gauge fixing term, the classical action \eqn{eq:classact} is invariant under the gauge transformation \eqn{eq:gauge}. This puts constraints on the $\varphi$- and $\cG$-dependence of the 2PI effective action $\Gammatpi[\varphi,\cG]$ and, in turn, on the $\varphi$-dependence of the 2PI-resummed effective action $\Gamma[\varphi]$. Equivalently, this results in nontrivial relations between various (2PI and/or 2PI-resummed) $n$-point vertex functions. To analyze this, we write the classical action as    
\beq
\label{eq:symaction}
 S[\varphi]=S_{\rm sym}[\varphi]+S_{\rm gf}[\varphi]
\eeq
where $S_{\rm sym}[\varphi]$ is the gauge-invariant classical QED action and $S_{\rm gf}[\varphi]$ is the gauge fixing term. 
We consider the infinitesimal transformation
\beq
\label{eq:inftransfo}
 \varphi\to\varphi+\delta^{(\alpha)}\varphi\,,
\eeq
where for general linearly realized symmetries\footnote{In this paper, we restrict our attention to the case where
bosonic and fermionic components of the superfield are not mixed by the symmetry transformation. This means that $\delta{\cal
A}_{mn}\neq0$ iff $(-1)^{q_m}=(-1)^{q_n}$ or, equivalently, $|q_m|=|q_n|$. In particular, the only nonvanishing
components of the matrix $\delta{\cal A}$ are $c$-numbers.}
\beq
\label{eq:lineartransfo}
 \delta^{(\alpha)}\varphi=\delta{\cal A}\,\varphi+\delta{\cal B}\,,
\eeq
with $\delta{\cal A}$ and $\delta{\cal B}$ field-independent two- and one-point functions of space-time respectively. For the $U(1)$ transformation \eqn{eq:gauge}, one has explicitly 
\beq
 \delta {\cal A}(x,y)=\delta^{(4)}(x-y)\,i\alpha(x) \,Q\,, 
\eeq
with $Q$ a diagonal matrix in superfield space, whose components are given by the fermionic charges: $Q_{mn}=q_m\delta_{mn}$, and 
\beq
\label{eq:B}
 \delta{\cal B}(x)=(-\p\alpha(x)/e,0,0)^t\,.
\eeq
Performing the change of variable $\hat\varphi\to\hat\varphi+\delta^{(\alpha)}\hat\varphi$, with $\delta^{(\alpha)}\hat\varphi\equiv\delta{\cal A}\,\hat\varphi+\delta{\cal B}$ in the path integral \eqn{eq:genfunc} and using the fact that the measure ${\cal D}\hat\varphi$ is invariant up to a multiplicative constant, one easily obtains, using standard manipulations, the following identity, which expresses the fact that the functional ${\cal W}[{\cal J},{\cal K}]$ is unaffected by this change of integration variables, at linear order in the infinitesimal transformation:
\beq
\label{eq:symid}
 \Big<\delta^{(\alpha)}S_{\rm gf}[\hat\varphi]+\delta^{(\alpha)}\hat\varphi_m\,{\cal
 J}_m+\delta^{(\alpha)}\hat\varphi_m\,\hat\varphi_n\,{\cal K}_{nm}\Big>=0\,,
\eeq
where 
\beq
 \delta^{(\alpha)}S_{\rm gf}[\hat\varphi]\equiv \delta^{(\alpha)}\hat\varphi_p\frac{\delta S_{\rm
 gf}[\hat\varphi]}{\delta\hat\varphi_p}
\eeq 
is the variation of the gauge-fixing action under \eqn{eq:inftransfo} at linear order in $\delta^{(\alpha)}\hat\varphi$.

Using the relations \eqn{eq:source1} and \eqn{eq:source2}, \Eqn{eq:symid} can be traded for a constraint equation for the 2PI functional:
\beq
\label{eq:2PIWI}
  \delta^{(\alpha)}\varphi_p\,\frac{\delta\Gamma_{\rm
  2PI}}{\delta\varphi_p}+\delta^{(\alpha)}\cG_{mn}\,\frac{\delta\Gamma_{\rm
  2PI}}{\delta\cG_{mn}}=\Big<\delta^{(\alpha)}S_{\rm gf}[\hat\varphi]\Big>\,,
\eeq
with $\delta^{(\alpha)}\varphi$ given by \Eqn{eq:lineartransfo} and
\beq
\label{eq:transfoGp}
 \delta^{(\alpha)}\cG\equiv
 \left<\Big(\delta^{(\alpha)}\hat\varphi\,\hat\varphi^t+\hat\varphi\,\delta^{(\alpha)}\hat\varphi^t\Big)\right>_{\!\!c}=\delta{\cal
 A}\,\cG+\cG\,\delta{\cal A}^t\,.
\eeq
Note that, here, the gauge variations $\delta^{(\alpha)}\varphi$ and $\delta^{(\alpha)}\cG$ have simple expressions because the symmetry under consideration is linear. Nonlinearly realized symmetries are discussed in App.~\ref{appsec:NPI}.

For linear gauges, the gauge fixing term is a quadratic functional of the electromagnetic field only. Since the latter has a purely affine gauge transformation, one has the obvious relation:
\beq
\label{eq:gf}
 \Big<\delta^{(\alpha)}S_{\rm gf}[\hat\varphi]\Big>=\delta^{(\alpha)}S_{\rm
 gf}[\varphi]=\delta^{(\alpha)}\varphi_p\frac{\delta S_{\rm gf}[\varphi]}{\delta\varphi_p}\,.
\eeq
Then Eq.~(\ref{eq:2PIWI}) can be rewritten as\beq
\label{eq:2PIWIlg}
  \left[\delta^{(\alpha)}\varphi_p\,\frac{\delta}{\delta\varphi_p}+\delta^{(\alpha)}\cG_{mn}\,\frac{\delta}{\delta\cG_{mn}}\right]\Big(\Gamma_{\rm
  2PI}[\varphi,\cG]-S_{\rm gf}[\varphi]\Big)=0\,,
\eeq
which states that the functional $\Gamma_{\rm 2PI}^{\rm sym}[\varphi,\cG]$ defined as
\beq
\label{eq:2PIWIsol}
 \Gamma_{\rm 2PI}^{\rm sym}[\varphi,\cG]\equiv\Gamma_{\rm 2PI}[\varphi,\cG]-S_{\rm gf}[\varphi]
\eeq
is invariant under
the infinitesimal (gauge) transformation defined by Eqs.~\eqn{eq:lineartransfo} and \eqn{eq:transfoGp}. \Eqn{eq:2PIWIlg} reflects the underlying (gauge) symmetry of the classical action at the level of the 2PI effective action. This generalizes the standard result that the gauge fixing term -- or any quadratic functional of the electromagnetic field only -- is not modified by loop (quantum) corrections. This result can easily be generalized to arbitrary higher ($n$PI) effective actions (see App.~\ref{appsec:NPI}).

Using \Eqn{eq:stat}, one recovers the standard Ward identities for the 2PI-resummed effective action
\eqn{eq:1PI}:\footnote{It is interesting to note that for theories with cubic interactions, \Eqn{eq:sclassint}, the 1PI
Ward identities can be entirely expressed in terms of the function $\bcG[\varphi]$. Indeed, inserting \Eqn{eq:firstder}
into \Eqn{eq:standardWI} and using the fact that, apart from the gauge fixing term, the classical action is
gauge-invariant, i.e. $\delta^{(\alpha)}\varphi_p\delta(S-S_{\rm gf})/\delta\varphi_p=0$, the 1PI Ward identities take the remarkably compact form:
$$
 \lambda_{pmn}\,\delta^{(\!\alpha\!)}\varphi_p\,\bcG_{mn}[\varphi]=0\,.
$$
For QED, this reads, explicitly:
$$
 -\frac{1}{e}\tr(\slashchar{\p}_x\bcG_{\psi\bar\psi}(x,x;\varphi))+i\slashchar{\bcG}_{A\bar\psi}(x,x;\varphi)\,\psi(x)-i\bar\psi(x)\,\slashchar{\bcG}_{\psi A}(x,x;\varphi)=0\,.
$$
where $(\slashchar{\bcG}_{A\bar\psi})_\alpha\equiv \bcG_{A\bar\psi}^{\mu\bar\alpha}\,\gamma_{\mu,\bar\alpha\alpha}$ and $(\slashchar{\bcG}_{\psi A})_{\bar\alpha}\equiv \gamma_{\mu,\bar\alpha\alpha}\,\bcG_{\psi A}^{\alpha\mu}$. We leave to the reader the instructive exercise to check that the above equation indeed generates standard Ward identities for 2PI-resummed vertex functions.
}
\beq
\label{eq:standardWI}
 \delta^{(\alpha)}\varphi_p\frac{\delta}{\delta\varphi_p}\Big(\Gamma[\varphi]- S_{\rm gf}[\varphi]\Big)=0\,,
\eeq
which states that the functional $\Gamma_{\rm sym}[\varphi]$ defined as
\beq
\label{eq:1PIWIsol}
 \Gamma_{\rm sym}[\varphi]\equiv\Gamma[\varphi]-S_{\rm gf}[\varphi]
\eeq
is invariant under the gauge-transformation \eqn{eq:lineartransfo}. One has clearly 
\beq
 \Gamma_{\rm sym}[\varphi]=\Gammatpi^{\rm sym}[\varphi,\bcG[\varphi]]\,.
\eeq

The standard way to exploit \Eqn{eq:standardWI} is to take successive derivatives of both sides of the equality with
respect to $\varphi$ and to set $\smash{\varphi=\bar\varphi}$. This directly generates the standard hierarchy of Ward identities
for the 2PI-resummed $n$-point functions \eqn{eq:npoint1}, see \Sec{sec:WI2PIres} below. One may be tempted to adopt a
similar strategy for exploiting \Eqn{eq:2PIWIlg}, by taking functional derivatives with respect to either $\varphi$ or
$\cG$. It can be checked by direct inspection that taking derivatives of \Eqn{eq:2PIWIlg} with respect to $\varphi$ does
not bring nontrivial information for cubic theories \eqn{eq:sclassint}. This is due to the fact that, in that case, the exact field dependence of $\Gammatpi[\varphi,\cG]$ is rather simple, see \Eqn{eq:sfielddep}. Differentiating \Eqn{eq:2PIWIlg} with respect to the propagator $\cG$ and setting $\smash{\cG=\bcG[\varphi]}$ generates nontrivial relations between so-called 2PI kernels, {\it i.e.} functions of the type $\delta^n\Gammaint/\delta\cG^n|_{\bcG[\varphi]}$ for arbitrary $\varphi$. However, it is not easy to extract useful information for, say the 2PI proper vertex functions \eqn{eq:npoint2}, from these relations. We shall not explore this direction further in the present paper.

In the next subsection, we follow a different strategy and exploit another aspect of the 2PI Ward identities
\eqn{eq:2PIWIlg}. We point out that the latter does not only restrict the functional form of the 2PI functional, but also
strongly constrains the field dependence of the correlator $\bcG[\varphi]$ or, equivalently, of the self-energy $\bar\Sigma[\varphi]$, see \Eqn{eq:EOM}. In turn, this gives direct constraints for the 2PI vertex functions \eqn{eq:npoint2}.

\subsection{Ward identities for 2PI vertex functions}

As an immediate consequence of \Eqn{eq:2PIWIsol}, one has that the physical correlator $\bar\cG[\varphi]$, defined as in
Eq.~(\ref{eq:stat}), is alternatively obtained as
\beq\label{eq:statsym}
\left.\frac{\delta\Gammatpi^{\rm sym}[\varphi,\cG]}{\delta\cG}\right|_{\bar\cG[\varphi]}=0\,,
\eeq
i.e. as the stationary point of a symmetric --
gauge-invariant in the sense defined in the previous section -- functional. Let us now show that this constrains the
field dependence of $\bar\cG[\varphi]$. Taking a derivative of Eq.~(\ref{eq:2PIWIlg}) with respect to $\cG$, one obtains
\beq\label{eq:eq}
\delta^{(\alpha)} \varphi_p\,\frac{\delta^2\Gamma^{\rm sym}_{\rm 2PI}}{\delta\varphi_p\delta\cG_{rs}}+\delta^{(\alpha)}\cG_{mn}\,\frac{\delta^2\Gamma_{\rm 2PI}^{\rm sym}}{\delta\cG_{mn}\delta \cG_{rs}}+\frac{\delta(\delta^{(\alpha)}\cG_{mn})}{\delta\cG_{rs}}\,\frac{\delta\Gamma_{\rm 2PI}^{\rm sym}}{\delta\cG_{mn}}=0\,.
\eeq
Now, from Eq.~(\ref{eq:statsym}) it follows that
\beq
\left.\frac{\delta^2\Gamma_{\rm 2PI}^{\rm sym}}{\delta \varphi_p\delta \cG_{rs}}\right|_{\bar\cG[\varphi]}+\left.\frac{\delta
\bar\cG_{mn}[\varphi]}{\delta\varphi_p}\frac{\delta^2\Gamma_{\rm 2PI}^{\rm sym}}{\delta\cG_{mn}\delta
\cG_{rs}}\right|_{\bar\cG[\varphi]}=0\,.
\eeq
Then, evaluating Eq.~(\ref{eq:eq}) for $\cG=\bar\cG[\varphi]$, one obtains
\beq
\left[\delta^{(\alpha)}
\varphi_p\frac{\delta\bar\cG_{mn}[\varphi]}{\delta\varphi_p}-\delta^{(\alpha)}\bar\cG_{mn}[\varphi]\right]\left.\frac{\delta^2\Gamma_{\rm
2PI}^{\rm sym}}{\delta\cG_{mn}\delta \cG_{rs}}\right|_{\bar\cG[\varphi]}=0\,.
\eeq
Assuming the invertibility of $\left.\delta^2\Gamma_{\rm 2PI}^{\rm sym}/\delta\cG\delta\cG\right|_{\bar\cG[\varphi]}$, one gets
\beq
\label{eq:WI}
\delta^{(\alpha)}\varphi_p\,\frac{\delta\bcG_{mn}[\varphi]}{\delta\varphi_p}=\delta^{(\alpha)}\bcG_{mn}[\varphi]\,.
\eeq
In other words, the function $\bcG[\varphi]$ transforms covariantly under an infinitesimal gauge transformation of its argument.\footnote{This can easily be understood as follows: By definition, the function $\bar\cG[\varphi]$ realizes the minimum of $\Gamma_{\rm 2PI}^{\rm
sym}[\varphi,\cG]$ as one varies $\cG$. Now gauge invariance reads, for infinitesimal transformations, 
$$
 \Gamma_{\rm 2PI}^{\rm sym}[\varphi,\bar\cG[\varphi]]=\Gamma_{\rm 2PI}^{\rm sym}[\varphi+\delta^{(\alpha)}\varphi,\bar\cG[\varphi]+\delta^{(\alpha)}\bar\cG[\varphi]]\,.
$$ It follows that $\bcG[\varphi]+\delta^{(\alpha)}\bcG[\varphi]$ realizes the minimum of $\Gamma_{\rm 2PI}^{\rm sym}[\varphi+\delta^{(\alpha)}\varphi,\cG]$ as one varies $\cG$. One then needs to have
$$ \bcG[\varphi+\delta^{(\alpha)}\varphi]=\bcG[\varphi]+\delta^{(\alpha)}\bcG[\varphi]\,,
$$
which is nothing but Eq.~(\ref{eq:WI}).}

Using \Eqn{eq:EOM} and  the fact that, by definition, the inverse free propagator $\cG^{-1}_{0}$ does not depend on the field $\varphi$, \Eqn{eq:WI} can be rewritten as
\beq
\label{eq:2PIWI2}
 \delta^{(\alpha)}\varphi_p\,\frac{\delta\bar\Sigma_{mn}[\varphi]}{\delta\varphi_p}=-\delta^{(\alpha)}\bcG^{-1}_{mn}[\varphi]\,,
\eeq
where 
\beq
\label{eq:transfobGpinv}
 \delta^{(\alpha)}\bcG^{-1}[\varphi]=-\delta{\cal A}^t\,\bcG^{-1}[\varphi]-\bcG^{-1}[\varphi]\,\delta{\cal A}\,. 
\eeq
\Eqn{eq:2PIWI2} is valid for arbitrary field $\varphi$. Taking functional derivatives with respect to the field $\varphi$ and setting $\smash{\varphi=\bar\varphi}$, one directly obtains a hierarchy of symmetry identities for the 2PI $n$-point vertex functions \eqn{eq:npoint2}. 
It is remarkable that these identities have the very same form as the usual Ward identities for the  1PI, or
2PI-resummed vertex functions \eqn{eq:npoint1}, as we show in the following. This is one of the main results of the
present paper. It is to be emphasized that this does not rely on the identification \eqn{eq:vertexidentity} of 2PI
vertex functions with 2PI-resummed vertex functions. We note that the analysis presented here trivially applies to the case of arbitrary (non-anomalous) linearly realized global symmetries (for which $S_{\rm gf}[\varphi]=0$). In that case, successive field-derivatives of \Eqn{eq:2PIWI2} generate the global form (i.e. involving an integration over space-time) of the corresponding Ward identities.\footnote{There are also local Ward identities associated with global symmetries, which involve correlation functions with insertion of the corresponding Noether current operator, see e.g. \cite{Pokorski}. These have been first discussed in the context of 2PI methods in Refs.~\cite{vanHees:2002bv}. Performing a similar analysis as presented here, one can derive the corresponding local identities for 2PI vertices. However, they are more complicated than the symmetry identities \Eqn{eq:2PIWI2} and we shall not discuss them further in this paper.}
 
Using Eqs.~\eqn{eq:lineartransfo}-\eqn{eq:B} and \eqn{eq:transfobGpinv}, we obtain the local version of the 2PI symmetry identity \eqn{eq:2PIWI2} in QED:
\bea
 &&\hspace{-1.cm}\Big[q_m\,\delta^{(4)}(x-z)+q_n\,\delta^{(4)}(z-y)\Big]i\bcG^{-1}_{mn}(x,y;\varphi)=\nn
\label{eq:2PIWI2comp}
 &&\frac{1}{e}\,\p^\mu_z\frac{\delta\bar\Sigma_{mn}(x,y;\varphi)}{\delta A^\mu(z)}+i\psi(z)\,\frac{\delta\bar\Sigma_{mn}(x,y;\varphi)}{\delta\psi(z)}-i\bar\psi(z)\,\frac{\delta\bar\Sigma_{mn}(x,y;\varphi)}{\delta\bar\psi(z)}\,.\nn
\eea
Below, we work out a few illustrative examples of the 2PI Ward identities which derive from \Eqn{eq:2PIWI2comp}. The simplest one is obtained by 
directly setting $\varphi=\bar\varphi=0$ in \Eqn{eq:2PIWI2comp}. The ($\bar\psi,\psi$)-component of the resulting equation reads
\beq
\label{eq:2PIWIex1}
 -\frac{1}{e}\,\p^\mu_zV^{(2,1)}_{\bar\alpha\alpha;\mu}(x,y;z)=\Big[\delta^{(4)}(x-z)-\delta^{(4)}(z-y)\Big]\,\bar D^{-1}_{\bar\alpha\alpha}(x,y)\,,
\eeq
where we employed a similar notation as in \Eqn{eq:defvertex} for the 2PI three-point function: 
\beq
\label{eq:def2PIvertex}
 V^{(2,1)}_{\bar\alpha\alpha;\mu}\equiv V^{(3)}_{\bar\psi_{\bar\alpha}\psi_\alpha A^\mu}=-i\left.\frac{\delta\bar\Sigma_{\bar\psi\psi}^{\bar\alpha\alpha}}{\delta A^\mu}\right|_{\bar\varphi}\,.
\eeq
\Eqn{eq:2PIWIex1} has the same form as the usual Ward identity relating the three-point photon-fermion vertex function and the inverse fermion two-point function, here with 2PI vertex functions.\footnote{For space-time translation invariant situations (e.g. in the vacuum), \Eqn{eq:2PIWIex1} can be written in momentum space, where it has the more familiar form, leaving Dirac indices implicit,
$$
 -\frac{1}{e}\,k^\mu\,V^{(2,1)}_{\mu}(p;k)=i\bar D^{-1}(p+k)-i\bar D^{-1}(p)\,,
$$
where the Fourier transforms of the functions $V^{(2,1)}_{\mu}$ and $\bar D^{-1}$ are defined through the usual relations: $(2\pi)^4\delta^{(4)}(p+p'+k)V^{(2,1)}_{\mu}(p;k)\equiv\int_{x,y,z}{\rm e}^{ip\cdot x+ip'\cdot y+ik\cdot z}V^{(2,1)}_{\mu}(x,y;z)$ and $(2\pi)^4\delta^{(4)}(p+p')\bar D^{-1}(p)\equiv\int_{x,y}{\rm e}^{ip\cdot x+ip'\cdot y}\bar D^{-1}(x,y)$.}
Similarly, taking successive derivatives of the $(A,A)$-component of \Eqn{eq:2PIWI2comp} with respect to the electromagnetic field $A$ and setting $\smash{A=0}$, $\smash{\psi=0}$ and $\smash{\bar\psi=0}$, we get, using a similar notation,
\beq
\label{eq:2PIWIex2}
 \p^{\sigma_i}_{z_i}V^{(0,p+2)}_{\mu\nu;\sigma_1\cdots\sigma_i\cdots\sigma_p}(x,y;z_1,\cdots,z_i,\cdots,z_p)=0\,,\qquad 1\le i\le p\,.
\eeq
Again, this is the 2PI version of the familiar result that $n$-photon vertex functions are transverse in momentum space. 

Notice that \Eqn{eq:2PIWIex2} only holds for 2PI $n$-photon vertex functions with $\smash{n\ge3}$. In fact, 2PI Ward identities do
not impose any constraint on the photon two-point function $i\bar G^{-1}\equiv i\bcG^{-1}_{AA}[\bar\varphi]$
\cite{Reinosa:2006cm}. This is rooted in the fact that, contrarily to higher 2PI vertex functions, the latter is defined
as the solution of a stationarity condition, \Eqn{eq:stat}, and not as a field derivative of some functional. Indeed, in
general, symmetries only constrain the functional dependence of various quantities of interest, such as the
$\varphi$ and $\cG$-dependence of the 2PI effective action $\Gammatpi[\varphi,\cG]$, \Eqn{eq:2PIWIlg}, the
$\varphi$-dependence of the 2PI-resummed effective action $\Gamma[\varphi]$, \Eqn{eq:standardWI}, or of the correlator
$\bcG[\varphi]$, \Eqn{eq:WI}. Therefore, symmetry identities only concern the derivatives of these functionals.

As a general result, to be emphasized, the fact that the 2PI two-point function $i\bcG^{-1}\equiv i\bcG^{-1}[\bar\varphi]$ does not, in general, satisfy the standard (1PI) Ward identities at finite approximation order does not constitute a direct violation of the symmetry constraints of the theory. For instance, in QED, the 2PI photon polarization tensor  -- i.e. the photon self-energy $\bar\Sigma_{AA}[\bar\varphi]$ -- does not have to be transverse in momentum space at any finite approximation order. Similarly, in scalar theories with spontaneously broken symmetry, the 2PI two-point function is not constrained to have a Goldstone mode at any finite approximation order.

This is to be contrasted with the 2PI-resummed two-point function $\Gamma^{(2)}$ which, being defined as a geometrical object -- here the curvature of the 2PI-resummed effective action in $\varphi$-space, see  \Eqn{eq:npoint1} --, is constrained by Ward identities \cite{Berges:2004vw}, see below. Only in the exact theory, where the 2PI-resummed and 2PI two-point vertex functions are identical, does the latter satisfies the usual Ward identities.

Another related remark concerns the fact that, among the $p+2$ legs of the 2PI vertex functions \Eqn{eq:npoint2}, two do not correspond to $\varphi$-derivatives, but to the legs of the self-energy $\bar\Sigma[\varphi]$. It also follows from the above discussion that (2PI) Ward identities generally do not apply to these particular legs. For instance, the identity \eqn{eq:2PIWIex1} applies to the vertex $\smash{V^{(3)}_{\bar\psi\psi;A}=-i\delta\bar\Sigma_{\bar\psi\psi}/\delta A}$, not to $\smash{V^{(3)}_{A\bar\psi;\psi}=-i\delta\bar\Sigma_{\bar\psi A}/\delta \psi=-i\delta\bar\Sigma_{A\psi}/\delta \bar\psi=-V^{(3)}_{\psi A;\bar\psi}}$. Although they also represent the photon-fermion vertex, the latter are not constrained by the underlying symmetry.
Similarly, \Eqn{eq:2PIWIex2} for 2PI $p+2$-photon functions only holds for the $p$ legs corresponding to $\varphi$-derivatives.

\subsection{Ward identities for 2PI-resummed vertex functions}
\label{sec:WI2PIres}

As shown previously, the 2PI-resummed effective action \eqn{eq:1PI} satisfies the standard Ward identities, \Eqn{eq:standardWI}, as it should \cite{Berges:2004vw} (see also \cite{Aarts:2002dj,vanHees:2002bv,Alejandro}). By taking derivatives of \Eqn{eq:standardWI} with respect to $\varphi$ and setting $\smash{\varphi=\bar\varphi=0}$, one generates the standard symmetry relations between 2PI-resummed proper vertex functions \eqn{eq:npoint1}. To compare the structure of these identities with the ones derived in the previous subsection for 2PI vertex functions, it is useful to introduce the 2PI-resummed two-point function in presence of a nonvanishing field:
\beq
 \Gamma^{(2)}_{mn}[\varphi]\equiv\frac{\delta^2\Gamma[\varphi]}{\delta\varphi_n\delta\varphi_m}\,.
\eeq
Obviously, 2PI-resummed $n$-point functions with $n\ge2$ can be obtained from the latter by taking field derivatives evaluated at $\smash{\varphi=\bar\varphi}$: $\smash{\Gamma^{(p+2)}=\delta^p\Gamma^{(2)}/\delta\varphi^p|_{\bar\varphi}}$.
Taking two field derivatives of \Eqn{eq:standardWI} and using the fact that the gauge fixing action is a quadratic
functional of the
electromagnetic field only, one readily obtains the following equation:
\beq
\label{eq:jaiplusdebonnom}
 \delta^{(\alpha)}\varphi_p\,\frac{\delta\Gamma^{(2)}_{mn}[\varphi]}{\delta\varphi_p}=\delta^{(\alpha)}\Gamma^{(2)}_{mn}[\varphi]\,,
\eeq
where
\beq
\label{eq:nonvraimentplus}
 \delta^{(\alpha)}\Gamma^{(2)}[\varphi]=-\delta{\cal A}^t\,\Gamma^{(2)}[\varphi]-\Gamma^{(2)}[\varphi]\,\delta{\cal A}\,.
\eeq
\Eqn{eq:jaiplusdebonnom} merely states that the variation of the two-point function $\Gamma^{(2)}[\varphi]$ under a
gauge-transformation \eqn{eq:lineartransfo} of its argument is given by the gauge-transformation $\delta^{(\alpha)}\Gamma^{(2)}[\varphi]$ of $\Gamma^{(2)}[\varphi]$. Taking further field derivatives of \Eqn{eq:jaiplusdebonnom}, one generates the standard Ward identities involving 2PI-resummed $n$-point functions with $n\ge3$. 

Eqs.~\eqn{eq:jaiplusdebonnom}-\eqn{eq:nonvraimentplus} are to be compared to \eqn{eq:2PIWI2}-\eqn{eq:transfobGpinv}. Since 2PI-resummed and 2PI vertices are obtained as $\varphi$-derivatives of $\Gamma^{(2)}[\varphi]$ and $\bar\Sigma[\varphi]$ respectively, see Eqs.~\eqn{eq:npoint1} and \eqn{eq:npoint2}, it is clear that, as announced earlier, Ward identities derived from \Eqn{eq:jaiplusdebonnom} for the former and from \Eqn{eq:2PIWI2} for the latter have the very same structure. We emphasize once again that this result does not depend on identifying 2PI-resummed and 2PI vertex functions, but relies solely on their respective definitions involving field derivatives. This is important since 2PI-resummed and 2PI vertex functions differ in general at finite approximation order. The remarkable result is that, if the given approximation satisfies the symmetry requirement \eqn{eq:2PIWIlg}, the vertex functions defined in \Eqn{eq:npoint1} and \Eqn{eq:npoint2} satisfy the standard Ward identities independently. 

For completeness, we write \Eqn{eq:jaiplusdebonnom} in a similar form as \Eqn{eq:2PIWI2}:
\bea
 &&\hspace{-1.2cm}-\Big[q_m\,\delta^{(4)}(x-z)+q_n\,\delta^{(4)}(y-z)\Big]i\Gamma^{(2)}_{mn}(x,y;\varphi)=\nn
\label{eq:1PIWI2}
 &&\frac{1}{e}\p^x_\mu\frac{\delta\Gamma^{(2)}_{mn}(x,y;\varphi)}{\delta A_\mu(z)}+i\psi(z)\,\frac{\delta\Gamma^{(2)}_{mn}(x,y;\varphi)}{\delta\psi(z)}-i\bar\psi(z)\,\frac{\delta\Gamma^{(2)}_{mn}(x,y;\varphi)}{\delta\bar\psi(z)}\,.\nn
\eea
As an illustration, the $(\bar\psi,\psi)$-component of this equation evaluated at $\smash{\varphi=\bar\varphi=0}$ is just the famous relation between the 2PI-resummed three-point and two-point functions:
\beq
 \frac{1}{e}\p^\mu_z\Gamma^{(2,1)}_{\bar\alpha\alpha\mu}(x,y,z)=
i\Big[\delta^{(4)}(x-z)-\delta^{(4)}(z-y)\Big]\,\Gamma^{(2,0)}_{\bar\alpha\alpha}(x,y)\,.
\eeq
This is to be compared to \Eqn{eq:2PIWIex1} for the corresponding 2PI vertices.

As already mentioned, there is an important difference between Ward identities for 2PI-resummed versus 2PI two-point
vertex functions. Contrarily to the latter, $i\bcG^{-1}[\bar\varphi]$, the 2PI-resummed two-point function
$\smash{\Gamma^{(2)}\equiv \Gamma^{(2)}[\bar\varphi]}$, being defined as a geometrical object, is constrained by a (1PI) Ward
identity, obtained by deriving \Eqn{eq:standardWI} once with respect to $\varphi$ and setting $\smash{\varphi=\bar\varphi}$:
\beq
\label{eq:WI2point}
 \delta^{(\alpha)}\bar\varphi_n\Gamma^{(2)}_{mn}=\left.\delta^{(\alpha)}\bar\varphi_n\frac{\delta S_{\rm gf}[\varphi]}{\delta\varphi_n\delta\varphi_m}\right|_{\bar\varphi},
\eeq
where $\delta^{(\alpha)}\bar\varphi\equiv\delta{\cal A}\,\bar\varphi+\delta{\cal B}$. In the QED case, with $\smash{\bar\varphi=0}$, \Eqn{eq:WI2point} states that the 2PI-resummed two-photon function $\Gamma^{(0,2)}$ satisfies
\beq
 \p_x^\mu\Gamma^{(0,2)}_{\mu\nu}(x,y)=\p^x_\mu\frac{\delta^2 S_{\rm gf}[\varphi]}{\delta A_\mu(x)\delta A_\nu(y)}\,.
\eeq
For space-time translation invariant situations, this is the standard statement that the longitudinal part of the photon two-point function in $4$-momentum space does not receive any loop correction. The corresponding 2PI-resummed photon polarization tensor $\Pi_{\mu\nu}\equiv\delta^2(\Gamma[\varphi]-S_0[\varphi])/\delta A^\nu\delta A^\mu$ is transverse in momentum space \cite{Berges:2004vw}:\footnote{Similarly, for $O(N)$ scalar theories, the 2PI-resummed two-point function does have exact Goldstone modes in the broken phase \cite{Aarts:2002dj,vanHees:2002bv}.}
\beq
\label{eq:Pitransv}
 \p_x^\mu \Pi_{\mu\nu}(x,y)=0\,.
\eeq

Finally, it is interesting to underline the deep relation between Ward identities for 2PI and 2PI-resummed vertex functions. For instance, recall that the 2PI-resummed two-photon function is related to 2PI vertex functions through \Eqn{eq:twoptb}. It is crucial that the 2PI three-point vertex $V^{(2,1)}\propto\delta\bar\Sigma_{\bar\psi\psi}/\delta A$ satisfies the Ward identity \eqn{eq:2PIWIex1} for \Eqn{eq:Pitransv} to hold.

\section{Gauge-invariant approximation schemes}
\label{sec:gaugeinvapprox}

So far, we have obtained (gauge) symmetry identities for the exact 2PI and 2PI-resummed effective actions and $n$-point vertex functions. For practical purposes, it is desirable to find approximation schemes which preserve these identities.\footnote{In particular, this is important for the purpose of renormalization \cite{Reinosa:2006cm,QED2}.} To this aim, it is enough to preserve \Eqn{eq:2PIWIlg}, or equivalently \Eqn{eq:2PIWIsol}. For instance, this directly implies the covariance property \eqn{eq:WI} of the (approximated) function $\bcG[\varphi]$, from which Ward identities for 2PI and 2PI-resummed vertex functions derive. Here we discuss symmetry preserving approximation schemes in the context of the standard 2PI diagrammatic expansion \cite{Cornwall:1974vz}.

Consider a general finite linear transformation of the superfield \eqn{eq:sfield}:
\beq
\label{eq:gtr1}
 \varphi\to\varphi^{(\alpha)}\equiv {\cal A}\,\varphi+{\cal B}\,,
\eeq 
where ${\cal A}$ is a matrix and ${\cal B}$ a vector in superfield space, the finite analogs of the infinitesimal $\delta {\cal A}$ and $\delta{\cal B}$ in \Eqn{eq:lineartransfo}.\footnote{We assume the same restrictions on ${\cal A}$ as before for $\delta{\cal A}$.}
The corresponding transformation of the connected correlator \eqn{eq:scor} is linear:
\beq
\label{eq:gtr2} 
 \cG\to\cG^{(\alpha)}\equiv {\cal A}\,\cG\,{\cal A}^t\,.
\eeq
For the $U(1)$ transformation \eqn{eq:gauge}, one has, explicitly, 
\beq
\label{eq:u1}
 {\cal A}(x,y)=\delta^{(4)}(x-y)\,\exp\Big(i\alpha(x) Q\Big)\,,
\eeq 
with $Q_{mn}=q_m\delta_{mn}$, and 
\beq
 {\cal B}(x)=(-\p\alpha(x)/e,0,0)^t\,.
\eeq 
For instance, the components of the gauge-transformed correlator $\cG^{(\alpha)}$ are given by
\beq
\label{eq:gtexpl}
 \cG^{(\alpha)}_{mn}(x,y)={\rm e}^{iq_m\alpha(x)}\,\cG_{mn}(x,y)\,{\rm e}^{iq_n\alpha(y)}\,.
\eeq
Notice, in particular, that, since that electromagnetic field $A$ has a purely affine transformation, the corresponding connected correlator $G=\cG_{AA}$ is invariant \cite{Reinosa:2006cm}:
\beq
\label{eq:photonpropgi}
 G^{(\alpha)}_{\mu\nu}(x,y)=G_{\mu\nu}(x,y)\,.
\eeq
This is another way to see that the underlying gauge symmetry cannot put any constraint on the 2PI photon two-point function.

The standard diagrammatic expansion of $\Gamma_{\rm 2PI}[\varphi,\cG]$ can be constructed from the so-called Cornwall-Jackiw-Tomboulis (CJT) parametrization of the 2PI effective action \cite{Cornwall:1974vz}. For the functional $\Gamma_{\rm 2PI}^{\rm sym}[\varphi,\cG]$ defined in Eq.~(\ref{eq:2PIWIsol}) it reads
\beq
\label{eq:CJTsym}
 \Gammatpi^{\rm sym}[\varphi,\cG]=S_{\rm sym}[\varphi]+\frac{i}{2}\,\Str\,\Ln\,\cG^{-1}+\frac{i}{2}\,\Str\,\cG_{\rm cl}^{-1}[\varphi]\,\cG+\Gamma_2[\varphi,\cG]\,,
\eeq
with the classical propagator
\beq
\label{eq:classprop}
 i\cG_{{\rm cl},mn}^{-1}[\varphi]\equiv(-1)^{q_n}\frac{\delta^{2} S[\varphi]}{\delta\varphi_m\delta\varphi_n}\,.
\eeq
and where $S_{\rm sym}[\varphi]$ has been defined in Eq.~(\ref{eq:symaction}). Here $\Gamma_2[\varphi,\cG]$ is the sum of closed 2PI diagrams with more than two-loops.\footnote{The CJT representation can be simply related to the interaction representation \eqn{eq:2PI} through
$$
 \Gammaint[\varphi,\cG]=S_{\rm int}[\varphi]+\frac{i}{2}\,\Str\,\cG_{\rm int}^{-1}[\varphi]\,\cG+\Gamma_2[\varphi,\cG]\,,
$$
where $i\cG_{{\rm int},mn}^{-1}[\varphi]\equiv(-1)^{q_n}\delta^{2} S_{\rm int}[\varphi]/\delta\varphi_m\delta\varphi_n$. As mentioned previously, for cubic theories, \Eqn{eq:sclassint}, the $\varphi$ dependence of the 2PI effective action is all contained in the classical and one-loop contributions and, therefore, the functional $\Gamma_2$ in \Eqn{eq:CJTsym} is $\varphi$-independent: $\Gamma_2[\varphi,\cG]\equiv\Gamma_2[\cG]$, see \Eqn{eq:sfielddep}.
} 

We would like to show that, within Eq.~(\ref{eq:CJTsym}), one can identify elementary contributions (each of which involves a finite number of terms) which are independently invariant under the transformation (\ref{eq:gtr1})-(\ref{eq:gtr2}) and thus which satisfy Eq.~(\ref{eq:2PIWIlg}). The first term in \Eqn{eq:CJTsym} is clearly gauge-invariant since, by definition,
\beq\label{eq:classgaugeinv}
S_{\rm sym}[\varphi]=S_{\rm sym}[\varphi^{(\alpha)}]\,.
\eeq
 As for the second term, it can be rewritten as a Gaussian functional integral (see App.~\ref{appsec:gaussian}):
\beq
 \frac{i}{2}\,\Str\,\Ln\,\cG^{-1}=-i\Ln\int{\cal D}\hat\varphi\,\exp\left\{\frac{1}{2}\,\hat\varphi^t\,i\cG^{-1}\,\hat\varphi\right\}+
 {\rm const.}\,,
\eeq
where `${\rm const.}$' denotes a $\cG$-independent contribution. A gauge transformation \eqn{eq:gtr2} of $\cG$, can be absorbed in a change of integration variable on the RHS. The corresponding Jacobian is simply a constant for the linear transformation \eqn{eq:gtr2}. Thus the second term on the RHS of \Eqn{eq:CJTsym} is gauge-invariant up to an additive, unphysical constant\footnote{For the $U(1)$ transformation \eqn{eq:u1}, one has $\Str\,\Ln{\cal A}=0$.}:
\beq
 {i\over 2}\Str\,\Ln\,\left(\cG^{(\alpha)}\right)^{-1}=\,{i\over 2}\Str\,\Ln\,\cG^{-1}-i\Str\,\Ln{\cal A}\,.
\eeq

To analyze the gauge transformation of the third (one-loop) term in \Eqn{eq:CJTsym}, we make use of the decomposition \eqn{eq:symaction} and write 
\beq
\label{eq:symgf}
 \cG_{\rm cl}^{-1}[\varphi]=\cG_{\rm sym}^{-1}[\varphi]+\cG_{\rm gf}^{-1}\,,
\eeq
where $\cG_{\rm sym}^{-1}[\varphi]$ and $\cG_{\rm gf}^{-1}$ are defined as $\cG_{\rm cl}^{-1}[\varphi]$ in
\Eqn{eq:classprop} with $S[\varphi]$ replaced respectively by $S_{\rm sym}[\varphi]$ and $S_{\rm gf}[\varphi]$. Note that for quadratic gauge fixing actions, $\cG_{\rm gf}^{-1}$ is independent of the field $\varphi$. It follows from \Eqn{eq:classgaugeinv} that
\beq
 {\cal A}^t\,\cG_{{\rm sym}}^{-1}[\varphi^{(\alpha)}]\,{\cal A}=\cG_{{\rm sym}}^{-1}[\varphi]\,,
\eeq
which immediately implies, using the cyclicity of the supertrace\footnote{One has: $
{\rm Str} (AB)=\sum_{n,m}(-1)^{q_n} A_{nm}B_{mn}
 =\sum_{n,m}(-1)^{q_m}B_{mn}A_{nm}
 ={\rm Str}(BA)$.}, that
\beq
 \Str\, \Big\{\cG_{\rm sym}^{-1}[\varphi^{(\alpha)}]\,\cG^{(\alpha)}\Big\}=\Str\,\Big\{\cG_{\rm sym}^{-1}[\varphi]\,\cG\Big\}\,.
\eeq
As noted above, the term $\cG_{\rm gf}^{-1}$ in \Eqn{eq:symgf} is independent of $\varphi$ and, therefore, invariant under the gauge transformation \eqn{eq:gtr1}. Moreover, since the gauge fixing action $S_{\rm gf}[\varphi]$ only depends on the electromagnetic field $A$, one has
\beq
 \Str\, \Big\{i\cG_{\rm gf}^{-1}\,\cG\Big\}=\Tr\,\Big\{\frac{\delta^{2} S_{\rm gf}}{\delta A\delta A}\,G\Big\}\,.
\eeq
This is clearly gauge-invariant since $G\equiv\cG_{AA}$ is, see \Eqn{eq:photonpropgi}, and, for linear gauges, $\delta^{2} S_{\rm gf}/\delta A\delta A$ is $A$-independent. 

Let us now consider the CJT functional $\Gamma_2[\varphi,\cG]$, which contains all the information beyond one-loop order. A given diagram contributing to $\Gamma_2[\varphi,\cG]$ is in general not invariant under \eqn{eq:gtr1}-\eqn{eq:gtr2} and one needs to consider particular subsets of diagrams to build gauge-invariant contributions. In particular, the vertices needed to construct the diagrammatic representation of $\Gamma_2[\varphi,\cG]$ are obtained from the shifted action $S[\varphi+\hat\varphi]$, as explained before. Some of these vertices depend on the field $\varphi$ and have, therefore, a non trivial gauge-transformation. For instance, it is clear that the affine contribution (i.e. ${\cal B}$ in \Eqn{eq:gtr1}) to the gauge-transformation of a $\varphi$-dependent vertex in a given diagram can only be canceled by the gauge-transformation of another diagram.

The construction of (gauge-)invariant subset of diagrams is made simple if one expresses the diagrammatic expansion of the $\Gamma_2[\varphi,\cG]$ in terms of the classical vertices
\beq
\label{eq:classvertices}
 \lambda^{(n)}_{1\cdots n}[\varphi]\equiv\frac{\delta^n S[\varphi]}{\delta\varphi_n\cdots\delta\varphi_1}\,.
\eeq
For $n\ge3$, there is no contribution from the (quadratic) gauge-fixing action to these vertices and, therefore, their gauge-transformation is completely determined by the symmetry property \eqn{eq:classgaugeinv}:
\beq
\label{eq:classvertexWI}
 \lambda^{(n)}_{m_1\cdots m_n}[\varphi^{(\alpha)}]=\lambda^{(n)}_{p_1\cdots p_n}[\varphi]\,{\cal
 A}_{p_1m_1}^{-1}\cdots{\cal A}_{p_nm_n}^{-1}\,.
\eeq
Notice that the classical vertices can contain differential operators which act on the local gauge-transformation matrices ${\cal A}^{-1}$ in \Eqn{eq:classvertexWI}. 
Now, consider such a vertex plugged into a given diagram in the 2PI expansion. There will be $n$ propagator lines attached to it, giving rise to a contribution of the form:
\beq
 \lambda^{(n)}_{m_1\cdots m_n}[\varphi]\,\cG_{m_1a_1}\cdots\cG_{m_na_n}\,,
\eeq
where the endpoints $a_1\cdots a_n$ are to be attached to other classical vertices, elsewhere in the diagram. A gauge transformation of this contribution reads
\bea
 &&\lambda^{(n)}_{m_1\cdots m_n}[\varphi^{(\alpha)}]\,\cG^{(\alpha)}_{m_1a_1}\cdots\cG^{(\alpha)}_{m_na_n}\nn
 &&\quad=
 \lambda^{(n)}_{m_1\cdots m_n}[\varphi^{(\alpha)}]\,{\cal A}_{m_1p_1}\cdots{\cal A}_{m_np_n}\,\cG_{p_1b_1}\cdots\cG_{p_nb_n}\,{\cal A}^t_{b_1a_1}\cdots{\cal A}^t_{b_na_n}\nn
 &&\quad=\lambda^{(n)}_{p_1\cdots p_n}[\varphi]\,\cG_{p_1b_1}\cdots\cG_{p_nb_n}\,{\cal A}^t_{b_1a_1}\cdots{\cal A}^t_{b_na_n}\,,
\eea
where we used \Eqn{eq:classvertexWI}.
Clearly, the gauge-transformations corresponding to the endpoints $m_1\cdots m_n$ of the propagators attached to the (classical) vertex under consideration cancel with the gauge-transformation of the vertex itself. The gauge-transformation corresponding to the endpoints $a_1\cdots a_n$ are to be canceled in a similar manner with that of the vertices to which these endpoints are attached in the (closed) diagram.\footnote{Notice that the one-loop term discussed previously is a particular case of the present argument, suitably adapted to the case for $n=2$.} In conclusion, we see that in order to construct (gauge-)invariant subsets of closed 2PI diagrams contributing to $\Gamma_2[\varphi,\cG]$, it is sufficient to organize the diagrammatic expansion in terms of the classical vertices \eqn{eq:classvertices}, which merely correspond to summing all vertices of the shifted action $S[\varphi+\hat\varphi]$ having the same number of $\hat\varphi$-legs. Indeed, it is this sum, and not the individual vertices, which has a simple gauge-transformation property, \Eqn{eq:classvertexWI}.

The situation is particularly simple for cubic theories \eqn{eq:sclassint}, such as QED, since, in that case, there is a single, $\varphi$-independent vertex:\footnote{That is the reason why the CJT functional $\Gamma_2$ is independent of $\varphi$, see \Eqn{eq:sfielddep}.}
\beq
 \lambda^{(3)}_{mnp}=\lambda_{mnp}\,.
\eeq
Therefore, the functional $\Gamma_2[\cG]$ is simply the set of closed 2PI diagrams made  with lines $\cG$ and the classical vertex. It follows from the above discussion that any such diagram is gauge-invariant. It is interesting to see the above general argument at work in this particular case. At each vertex of the diagram are attached three lines, giving a term of the form
\beq
\label{eq:gaugecancel}
 \lambda_{mnp}\,\cG_{ma_1}(x,x_1)\,\cG_{na_2}(x,x_2)\,\cG_{pa_3}(x,x_3)\,,
\eeq
where $x$ is the space-time location of the vertex, to be integrated over, and the $a_i$'s and $x_i$'s are to be attached to other vertices in the diagram. In QED, the local phase factor from the gauge transformation \eqn{eq:gtexpl} associated with this vertex is given by ${\rm e}^{i(q_m+q_n+q_p)\alpha(x)}=1$ since the classical vertex $\lambda_{mnp}\neq0$ only for $\smash{q_m+q_n+q_p=0}$. In conclusion, any approximation of the 2PI effective action based on the 2PI diagrammatic expansion is gauge invariant.

In theories with more than cubic interactions, higher order classical vertices appear in the construction of $\Gamma_2[\varphi,\cG]$. In that case, individual diagrams are not separately gauge-invariant. The procedure described above allows one to build systematic gauge-invariant approximations. This is illustrated in App.~\ref{appsec:SQED} in the case of SQED.

\section{Conclusion and Outlook}

In this paper, we have studied the implications of the symmetries of the classical action on the quantum theory in the context of 2PI functional techniques\footnote{Higher ($n$PI) effective actions are discussed in App.~\ref{appsec:NPI}.} for abelian gauge theories. Our results apply to theories with arbitrary linear global symmetries as well. 

It is known that approximations based on 2PI techniques generally do not satisfy the standard Ward identities associated with the underlying symmetries of the theory. This fact should, however, be interpreted with care. The main point of the present paper is to emphasize that symmetry constraints on the 2PI effective action can be quite different from the usual ones for the 1PI effective action. A careful analysis of symmetry constraints in the 2PI framework in fact reveals that the relevant (i.e. 2PI) Ward identities are fulfilled by generic 2PI approximations.

We have derived generalized (2PI) Ward identities for the various $n$-point vertex functions in the 2PI framework and have devised a systematic procedure to build approximation schemes which satisfy the symmetry constraints at any finite order in the context of the diagrammatic representation of the 2PI functional. For instance, in QED, each closed 2PI diagram satisfies the required symmetry constraints. As a consequence, the 2PI Ward identities are exactly satisfied at any approximation order. 

Furthermore, we have shown that the inverse propagator $i\bcG^{-1}[\varphi]$, which defines the extremum of the 2PI effective action in propagator space, is not directly constrained by the underlying symmetry. Therefore, the well-known fact that the latter does not satisfy usual symmetry constraints (e.g. the corresponding photon polarization tensor is not transverse in momentum space) for generic approximations does not constitute a violation of (2PI) Ward identities. 

We have also shown that, as a consequence of 2PI Ward identities, the so-called 2PI-resummed effective action, i.e. the 2PI functional evaluated at its extremum in propagator space, $\Gamma[\varphi]\equiv\Gammatpi[\varphi,\bcG[\varphi]]$, satisfies the standard symmetry constraints, as expected \cite{Berges:2004vw}. This generalizes previous results for global symmetries \cite{Aarts:2002dj,vanHees:2002bv,Berges:2005hc} (see also \cite{Alejandro}) to abelian gauge symmetries. In other words, the 2PI-resummed vertex functions, obtained as derivatives of the 2PI-resummed effective action evaluated at the physical point $\varphi=\bar\varphi$, satisfy standard Ward identities.\footnote{We stress that mixed correlators play a crucial role for this to be true.} For instance, the 2PI-resummed photon polarization tensor in QED is transverse in momentum space at any approximation order.

We have identified another class of $n$-point functions in the 2PI framework which exactly satisfy the standard Ward identities at any approximation order. These ``2PI vertex functions'' are obtained as field-derivatives of the two-point function $\bcG^{-1}[\varphi]$ or, equivalently, of the self-energy $\bar\Sigma[\varphi]$, evaluated at the physical point $\varphi=\bar\varphi$. Alternatively, they can be obtained as solutions of appropriate Bethe-Salpeter--like equations. Although 2PI-resummed and 2PI vertex functions are identical in the exact theory, they differ in general at finite approximation order. It is remarkable that they separately exactly satisfy the standard Ward identities (except for the 2PI two-point function) at any order of approximation.

It is important to stress that the present results concern the bare (unrenormalized) theory. A crucial issue is to check that the symmetry constraints mentioned above are preserved after renormalization. This is the purpose of \Ref{QED2}. There we show that the Ward identities derived in the present paper play a crucial role in constraining the possible UV divergences of the various $n$-point functions in the 2PI framework. In turn, the counterterms needed to renormalize the theory at any finite approximation order do satisfy the symmetry constraints derived here \cite{QED2}. 

We believe the present analysis of 2PI Ward identities together with the renormalization proofs presented in Refs.~\cite{Reinosa:2006cm,QED2} provide a solid theoretical basis for practical applications of 2PI techniques to abelian gauge theories, see e.g. \cite{Borsanyi:2007bf}. The study of nonabelian gauge theories deserves further investigation.

\section*{Acknowledgments}
We thank J. Berges and Sz. Bors\'anyi for interesting discussions during the early stages of this work and A.
Arrizabalaga for letting us know about \Ref{Alejandro}. J.S. thanks C. Volpe and Luna for their support during the
redaction of this paper. U.R. thanks the Alexander von Humboldt foundation for support within a research fellowship.

\appendix

\section{Slavnov-Taylor and Ward identities for $n$PI effective actions}
\label{appsec:NPI}

We first recall the definitions of the $n$-particle-irreducible ($n$PI) effective action \cite{DeDom:1964vz} (see also \cite{Berges:2004pu}).
We consider a general field theory for a set of bosonic and/or fermionic fields which we collectively represent by a superfield $\varphi$. 
All correlation functions of the theory can be obtained from the generating functional 
\beq
\label{appeq:genfunc}
 {\rm e}^{iW[{\cal K}]}\equiv {\rm e}^{iW[{\cal K}^{(1)},\ldots,{\cal K}^{(n)}]}=\int{\cal D}\hat\varphi\,{\rm e}^{iS[\hat\varphi]+i\sum_{p=1}^n\frac{1}{p!}\,\hat\varphi^p\,{\cal K}^{(p)}}\,,
\eeq
where $S[\varphi]$ is the classical action of the theory and ${\cal K}^{(1)},\ldots,{\cal K}^{(n)}$ denote a set of $1$-,$\ldots$,$n$-point classical sources, coupled to products of $1,\cdots,n$ fields respectively. Here, we employ the shorthand notation:
\beq
\label{appeq:notation}
 \hat\varphi^p\,{\cal K}^{(p)}\equiv\hat\varphi_{n_1}\ldots\hat\varphi_{n_p}\,{\cal K}^{(p)}_{n_p\ldots n_1}\,.
\eeq
Notice that, due to the Grassmanian character of the fermionic components of $\varphi$, the sources have the following symmetry property under permutation of any neighboring indices:
\beq
\label{appeq:symsource}
 {\cal K}^{(p)}_{n_1 n_2\ldots n_p}=(-1)^{q_{n_1}q_{n_2}}{\cal K}^{(p)}_{n_2 n_1\ldots n_p}\,.
\eeq
Correlation functions of superfield operators can be obtained by differentiation with respect to the classical sources. For instance, one has
\beq
\label{appeq:correlators}
 \frac{\delta_L W[{\cal K}]}{\delta {\cal K}^{(p)}_{n_p\ldots
 n_1}}=\frac{1}{p!}\,\Big<\hat\varphi_{n_1}\ldots\hat\varphi_{n_p}\Big>\equiv \frac{1}{p!}\,\cC^{(p)}_{n_1\ldots n_p}\,,
\eeq
where the brackets denote an expectation value in presence of the classical sources, generalizing the definition \eqn{eq:expvalue} to the case of $p$-point sources with $2\le p\le n$.
Notice that the functions $\cC_{n_1\ldots n_p}$ defined in \Eqn{appeq:correlators} represent the full $p$-point correlators, including unconnected contributions. We shall denote by $\cG_{n_1\ldots n_p}$ the corresponding connected $p$-point correlators (or cumulants). One has, for instance, $\cG^{(1)}_1=\cC^{(1)}_1$ for the one-point correlator, $\cG^{(2)}_{12}=\cC^{(2)}_{12}-\cC^{(1)}_1\cC^{(1)}_2$ for the two-point correlator, etc. Using a shorthand notation similar to the one employed before, see \Eqn{appeq:notation}, \Eqn{appeq:correlators} can be written as
\beq
\label{appeq:correlators2}
 \frac{\delta_L W[{\cal K}]}{\delta {\cal K}^{(p)}}=\frac{1}{p!}\,\cC^{(p)}\,.
\eeq

The $n$-particle-irreducible ($n$PI) effective action is defined as the multiple Legendre transform of the generating functional \eqn{appeq:genfunc}:
\beq
\label{appeq:nPI}
 \Gamma_{\rm nPI}[\cG]\equiv\Gamma_{\rm nPI}[\cG^{(1)},\ldots,\cG^{(n)}]=W[{\cal K}]-\sum_{p=1}^n\frac{\delta_L W[{\cal K}]}{\delta {\cal K}^{(p)}}\,{\cal K}^{(p)}
\eeq
It is traditionally parametrized in terms of the connected correlators $\cG^{(p)}$. However, for our discussion on Slavnov-Taylor identities below, it proves useful to exploit its dependence on the full correlators, the $\cC$'s. That is the reason why, in the following, we often consider functional derivatives of the $n$PI effective action with respect to the $\cC$'s. For instance, using the same notation as before, one has the relations
\beq
\label{appeq:source}
 \frac{\delta_R \Gamma_{\rm nPI}[\cG]}{\delta \cC^{(p)}}\equiv\frac{\delta_R \Gamma_{\rm nPI}[\cG]}{\delta \cC^{(p)}_{n_p\ldots n_1}}=-\frac{1}{p!}\,{\cal K}^{(p)}_{n_1\ldots n_p}\equiv-\frac{1}{p!}\,{\cal K}^{(p)}\,.
\eeq
Physical correlation functions are obtained for vanishing external sources.
 
The (1PI) effective action of the theory $\Gamma[\varphi]$, that is the generating functional for $n$-point vertex functions can be obtained as the $n$PI functional \Eqn{appeq:nPI} for vanishing $p$-point sources with $2\le p\le n$. Writing $\varphi\equiv\cG^{(1)}$ for the one-point function, we have
\beq
\label{appeq:nPIresummed}
 \Gamma[\varphi]\equiv \Gamma_{\rm 1PI}[\varphi]=\Gamma_{\rm nPI}[\varphi,\bcG^{(2)}[\varphi],\cdots,\bcG^{(n)}[\varphi]]
\eeq
where the functions $\bcG^{(p)}[\varphi]$, are obtained from  Eqs.~(\ref{appeq:source}) at vanishing sources, which are easily shown to be equivalent to the stationarity conditions
\beq
\label{appeq:barG}
 \left.\frac{\delta_R\Gamma_{\rm nPI}[\cG]}{\delta\cG^{(p)}}\right|_{\bcG[\varphi]}=0\quad{\rm for}\quad 2\le p\le n\,.
\eeq
\Eqn{appeq:nPIresummed} defines the $n$PI-resummed effective action.

We now consider the constraints that symmetries of the classical theory put on $n$PI effective actions.
Here, we consider a general (local, nonlinear, etc.) continuous symmetry.
We write the classical action as
\beq
 S[\varphi]=S_{\rm sym}[\varphi]+S_{\rm sb}[\varphi]
\eeq
where $S_{\rm sym}[\varphi]$ is invariant under the infinitesimal transformation
\beq
\label{appeq:transfo}
 \varphi\to\varphi+\delta^{(\alpha)}\varphi\,,
\eeq
and we allow for an explicit symmetry breaking term $S_{\rm sb}[\varphi]$ (for instance a gauge-fixing term in gauge theories). In general, the infinitesimal variation $\delta^{(\alpha)}\varphi$ can be a nonlinear, space-time dependent, functional of $\varphi$.

Using similar manipulations as those leading to \Eqn{eq:symid}, one easily obtains the following symmetry identity:
\beq
\label{appeq:ident1}
 \left<\delta^{(\alpha)}S_{\rm sb}[\hat\varphi]+\sum_{p=1}^n\frac{1}{(p-1)!}\,\delta^{(\alpha)}\hat\varphi\,\hat\varphi^{p-1}\,{\cal K}^{(p)}\right>=0\,,
\eeq
where we used a similar notation as before:\footnote{In deriving \Eqn{appeq:ident1}, we made use of the symmetry property \Eqn{appeq:symsource}.}
\beq
 \delta^{(\alpha)}\hat\varphi\,\hat\varphi^{p-1}\,{\cal K}^{(p)}\equiv\delta^{(\alpha)}\hat\varphi_{n_1}\,\hat\varphi_{n_2}\ldots\hat\varphi_{n_p}\,{\cal K}^{(p)}_{n_p\ldots n_1}\,.
\eeq

Defining the variation of the correlators $\cC^{(p)}$, introduced previously, as
\beq 
\label{appeq:transfoGp}
 \delta^{(\alpha)}\cC^{(p)}_{n_1\ldots n_p}=
 \Big<\delta^{(\alpha)}\hat\varphi_{n_1}\,\hat\varphi_{n_2}\ldots\hat\varphi_{n_p}\Big>+\ldots+\Big<\hat\varphi_{n_1}\,\hat\varphi_{n_2}\ldots\delta^{(\alpha)}\hat\varphi_{n_p}\Big>\,,
\eeq
and using \Eqn{appeq:source} to eliminate the sources ${\cal K}^{(p)}$ in \Eqn{appeq:ident1}, one has
\beq
\label{appeq:nPIST1}
  \sum_{p=1}^n\delta^{(\alpha)}\cC^{(p)}\,\frac{\delta_R\Gamma_{\rm nPI}[\cG]}{\delta\cC^{(p)}}=\Big<\delta^{(\alpha)}S_{\rm sb}[\hat\varphi]\Big>\,,
\eeq
or, equivalently, recognizing the LHS as the change $\delta^{(\alpha)}\Gamma_{\rm nPI}[\cG]$ of the $n$PI functional under the symmetry transformation \Eqn{appeq:transfoGp},
\beq
\label{appeq:nPIST2}
 \delta^{(\alpha)}\Gamma_{\rm nPI}[\cG]=\sum_{p=1}^n\delta^{(\alpha)}\cG^{(p)}\,\frac{\delta_R\Gamma_{\rm nPI}[\cG]}{\delta\cG^{(p)}}=\Big<\delta^{(\alpha)}S_{\rm sb}[\hat\varphi]\Big>\,.
\eeq
Here the sources ${\cal K}$ on the RHS are given as functions of the $\cG$'s by \Eqn{appeq:source}.
\Eqn{appeq:nPIST2} makes clear that, although, we made use of the correlators $\cC^{(p)}$ in deriving \Eqn{appeq:nPIST1}, everything can finally be expressed in terms of the connected correlators $\cG^{(p)}$. The variations $\delta^{(\alpha)}\cG^{(p)}$ under the infinitesimal transformation can be simply obtained from \Eqn{appeq:transfoGp} by replacing the expectation values $\langle\cdots\rangle$ by connected expectation values $\langle\cdots\rangle_c$. 
For vanishing symmetry breaking $S_{\rm sb}[\varphi]=0$, \Eqn{appeq:nPIST2} states that the $n$PI functional is invariant under the transformations $\cG^{(p)}\to\cG^{(p)}+\delta^{(\alpha)}\cG^{(p)}$, as defined above. This generalizes the standard (1PI) Slavnov-Taylor identity to $n$PI effective actions. Of course \Eqn{appeq:nPIST2} contains, as a particular case for $n=1$, the standard (1PI) Slavnov-Taylor identities. Equivalently, setting $\cG^{(p)}\to\bcG^{(p)}[\varphi]$ for $2\le p\le n$ in \Eqn{appeq:nPIST1} and using \Eqn{appeq:barG}, one recovers, in the absence of symmetry breaking, the standard Slavnov-Taylor identities for the $n$PI-resummed effective action $\Gamma[\varphi]$ \cite{Weinberg}.

In general, Slavnov-Taylor identities are difficult to exploit because the transformations (\ref{appeq:transfoGp}) of the
$p$-point correlation functions involve higher-order correlation functions, which are not arguments of the $n$PI
functional and, therefore, need to be calculated for arbitrary sources. However, this is not so for the class of linear
symmetries considered in the present paper, where $F[\varphi]={\cal A}\,\varphi+{\cal B}$, for which the transformation (\ref{appeq:transfoGp}) of the $p$-point correlator only involves lower or equal order correlation functions.

The results of \Sec{sec:sym} are easily generalized to $n$PI effective actions. For instance for QED (or SQED) in linear gauges, the symmetry-breaking action is given by the gauge-fixing term: $S_{\rm sb}[\varphi]=S_{\rm gf}[\varphi]$, which is a quadratic functional of the electromagnetic field only. In that case, one easily shows that the functional
\beq
\label{appeq:nPIWI}
 \Gamma_{\rm nPI}^{\rm sym}[\cG]=\Gamma_{\rm nPI}[\cG]-S_{\rm gf}[\varphi]
\eeq
is gauge-invariant, i.e. invariant under the gauge-transformation \eqn{appeq:transfoGp}. Taking functional derivatives of this equation with respect to the $\cG^{(p)}$'s, one generates the $n$PI generalization of Ward identities.
Moreover, since the gauge-fixing (or explicit symmetry-breaking) term in \eqn{appeq:nPIWI} only depends on the one-point correlator
$\varphi\equiv\cG^{(1)}$, one has
\beq
 \frac{\delta_R\Gamma_{\rm nPI}}{\delta\cG^{(p)}}=\frac{\delta_R\Gamma_{\rm nPI}^{\rm sym}}{\delta\cG^{(p)}}\quad{\rm for}\quad p\ge2\,.
\eeq
It follows that, compare to \Eqn{eq:WI},
\beq\label{appeq:nPIWI2}
\delta^{(\alpha)}\varphi_q\frac{\delta_R\bcG^{(p)}[\varphi]}{\delta\varphi_q}= \delta^{(\alpha)}\bcG^{(p)}[\varphi]\quad{\rm for}\quad p\ge2\,,
\eeq
where the variations $\delta^{(\alpha)}\bcG^{(p)}[\varphi]$ are defined as in \Eqn{appeq:transfoGp} with ${\cal K}^{(p\ge2)}=0$. 
\Eqn{appeq:nPIWI2} is valid for arbitrary field $\varphi$. Taking functional derivatives of both sides of the equality
and setting $\varphi=\bar\varphi$, one generates an infinite hierarchy of $n$PI Ward identities relating various
$p$-point functions of the theory. For the exact theory, they are identical to the usual Ward identities. However, for
finite approximations, where the relation between the variational vertex functions $\bcG_p$ and the $p$-point functions obtained by functional derivatives of the ($n$PI-resummed) effective action $\Gamma[\varphi]$ is not satisfied in general, these identities are non-trivial.

\section{The 2PI effective action in the superfield formalism}
\label{appsec:gaussian}

Using the textbook formulae for multi-dimensional Gaussian integrals on $c$-numbers $x_i$, $i=1,\ldots,N_b$, and $a$-numbers $\xi_a$, $a=1,
\ldots,N_f$:
\beq
 \int d^{N_b}x\,\,{\rm e}^{-\frac{1}{2}x_ix_j M_{ji}+ x_i J_i}\propto{\rm e}^{-\frac{1}{2}\tr \ln M+ \frac{1}{2} x_i M^{-1}_{ij} x_j}\,,
\eeq
where $M$ is a symmetric $N_b\times N_b$ matrix and $J$ an $N_b$-component vector of $c$-numbers, and 
\beq
 \int d^{N_f}\xi\,\,{\rm e}^{-\frac{1}{2}\xi_a\xi_b H_{ba} + \xi_a\eta_a}\propto{\rm e}^{+\frac{1}{2}\tr \ln H+\frac{1}{2}\eta_a H^{-1}_{ab} \eta_b}\,,
\eeq 
where $H$ is an antisymmetric $N_f\times N_f$ matrix of $c$-numbers and $\eta$ is an $N_f$-component vector of $a$-numbers, it is a straightforward -- though nontrivial -- exercise to show that \cite{DeWitt,Calzetta:2004sh}
\beq
 \int d^N\chi\,\,{\rm e}^{-\frac{1}{2}\chi_m\chi_n {\cal M}_{nm} + \chi_m{\cal J}_m}\propto{\rm e}^{-\frac{1}{2}\str \ln {\cal M}+\frac{1}{2}{\cal J}_m{\cal M}^{-1}_{mn} {\cal J}_n}\,,
\eeq
where $\chi=(x,\xi)^t$ is an $N$-component superfield ($N=N_b+N_f$), $d^N\chi\equiv d^{N_b}x\,d^{N_f}\xi$ and $\str$ denotes the supertrace:
\beq
 \str\,{\cal M}\equiv\sum_m(-1)^{q_m}{\cal M}_{mm}\,.
\eeq 
Here, the $N\times N$ matrix ${\cal M}$ is such that ${\cal M}_{mn}=(-1)^{q_mq_n}{\cal M}_{nm}$. The component ${\cal M}_{mn}$ is a $c$-number if $q_m+q_n$ is even, an $a$-number otherwise. Similarly, the component ${\cal J}_m$ of the supervector ${\cal J}$ is a $c$-number if $q_m$ is even, an $a$-number otherwise.

For the free theory, with classical action given by \Eqn{eq:freeact}, the generating functional \eqn{eq:genfunc} is Gaussian and can be computed exactly (beware of factors~$i$):
\beq
 W[{\cal J},{\cal K}]=\frac{i}{2}\,\Str \ln {\cal M}+\frac{i}{2}\,{\cal J}_m{\cal M}^{-1}_{mn}\,{\cal J}_n\,,
\eeq
where $\Str$ denotes the functional supertrace (it involves a space-time integration) and where
\beq
 {\cal M}_{mn}\equiv(-1)^{q_n}\cG_{0,mn}^{-1}-i{\cal K}_{mn}\,.
\eeq
Using Eqs.~\eqn{eq:onepointdef}-\eqn{eq:twopointdef}, one easily obtains, for the one- and two-point connected correlators,
\beq
 \varphi_m=i{\cal J}_n{\cal M}^{-1}_{nm}
\qquad{\rm and}\qquad
 \cG_{mn}=(-1)^{q_n}{\cal M}^{-1}_{mn}\,.
\eeq
Finally, the Legendre transform \eqn{eq:legendre} is easily calculated. It can be expressed as
\beq
\label{appeq:free2PI} \Gammatpi[\varphi,\cG]=S_0[\varphi]+\frac{i}{2}\,\Str\ln\cG^{-1}+\frac{i}{2}\,\Str\,\cG_0^{-1}\cG+{\rm const.}
\eeq

This generalizes the standard expressions of the free part of 2PI functional in the cases where only bosonic or fermionic degrees of freedom are involved
\cite{Luttinger:1960ua,Cornwall:1974vz,DeDom:1964vz}.\footnote{The standard expression of the 2PI effective action in the case with only fermionic degrees of freedom \cite{Cornwall:1974vz}, discarding mixed correlators $F$ and $\bar F$ (see \Eqn{eq:cor3}), can be obtained from \Eqn{appeq:free2PI}, by replacing $\cG_0^{-1}$ and $\cG$ by the free inverse fermion propagator $D_0^{-1}$ and the connected correlator $D$ respectively, and by replacing the factors $i/2$ on the RHS by factors $-i$. The minus sign comes from the definition of the supertrace and the factor $2$ comes from the fact that for relativistic (Dirac) fermion species, one has to introduce a doublet of spinor fields $\xi\equiv(\psi,\bar\psi^t)^t$.} Notice that in the general case, with both bosonic and fermionic degrees of freedom, mixed components of the correlator must be included and the 2PI effective action is not just the sum of the purely bosonic and the fermionic parts.

For the theory with interactions, one can parametrize the total effective action as in \Eqn{eq:2PI}. The construction of the interaction part $\Gammaint[\varphi,\cG]$ as the set of closed 2PI diagrams with lines given by $\cG$ and vertices obtained from the shifted action $S[\varphi+\hat\varphi]$ follows the steps of \Ref{Cornwall:1974vz}.

\section{2PI and 2PI-resummed vertex functions}
\label{sec:appvertex}

Here, we show that 2PI-resummed and 2PI proper vertex functions, defined in Eqs.~\eqn{eq:npoint1} and \eqn{eq:npoint2}, are identical in the exact theory, see also \cite{Cornwall:1974vz}. 
We start with the identity:
\beq
 \delta\varphi_m=\frac{\delta_L\varphi_m}{\delta{\cal J}_p}\delta\varphi_n\frac{\delta_R{\cal J}_p}{\delta\varphi_n}\,,
\eeq
valid for an arbitrary variation $\delta\varphi$. It follows that
\beq
\label{appeq:id}
 (-1)^{q_n+q_nq_p}\frac{\delta_L\varphi_m}{\delta{\cal J}_p}\frac{\delta_R{\cal J}_p}{\delta\varphi_n}=\delta_{mn}\,.
\eeq
Now, it is clear from \Eqn{eq:genfunc} that, in the exact theory,
\beq
\label{appeq:djdj}
 \left.\frac{\delta_L\varphi_m}{\delta{\cal J}_n}\right|_{{\cal K}=0}=\frac{\delta_L^2W[{\cal J},{\cal
 K}=0]}{\delta{\cal J}_m\delta{\cal J}_n}=i\bcG_{mn}[\varphi]\,,
\eeq
where it is understood that the leftmost derivative (here $\delta_L/\delta{\cal J}_m$) is to be taken first. One has also, from Eqs.~\eqn{eq:source1}, \eqn{eq:stat} and \eqn{eq:1PI},
\beq
 \left.\frac{\delta_R{\cal J}_p}{\delta\varphi_n}\right|_{{\cal K}=0}=-\frac{\delta_R^2\Gamma[\varphi]}{\delta\varphi_n\delta\varphi_p}\,.
\eeq
Therefore, \Eqn{appeq:id} reads
\beq
 \bcG_{mp}[\varphi]\,\frac{\delta_R^2\Gamma[\varphi]}{\delta\varphi_p\delta\varphi_n}=i(-1)^{q_n}\delta_{mn}\,,
\eeq
which is equivalent to \Eqn{eq:identity}. The identity between 2PI-resummed and 2PI vertex functions in the exact theory, \Eqn{eq:vertexidentity}, follows.

\section{Gauge-invariant approximations in 2PI scalar QED}
\label{appsec:SQED}

The classical action for scalar QED (SQED) reads, in covariant gauge,
\bea
\label{appeq:SQED}
 S[A,\phi,\phi^\dagger]&=&\int_x \Big\{-\phi^\dagger(\partial^2+m^2)\phi+\frac{1}{2}A^\mu\Big[g_{\mu\nu}\partial^2-(1-\lambda)\partial_\mu\partial_{\nu}\Big]A^\nu\nn
 &&\qquad-ieA^\mu[\phi\,\partial_\mu\phi^\dagger-\phi^\dagger\partial_\mu\phi]+e^2 A^\mu A_\mu\phi^\dagger\phi\Big\}\,,
\eea
where $\phi$ is a complex scalar field and $\lambda$ is the gauge-fixing parameter. Apart from the gauge-fixing term, it is invariant under the gauge transformation
\beq\label{appeq:scalargauge}
\phi(x)\rightarrow e^{i\alpha(x)}\phi(x)\,,\;\;
\phi^\dagger(x)\rightarrow e^{-i\alpha(x)}\phi^\dagger(x)\,,\;\;
A_\mu(x)\rightarrow A_\mu(x)-\frac{1}{e}\,\partial_\mu\alpha(x)\,,\nonumber\\
\eeq
where $\alpha(x)$ is an arbitrary real function. Following the analysis presented in the paper,
one introduces the superfield $\varphi\equiv(A,\phi,\phi^\dagger)^t$ and the corresponding correlator $\cG$ as in \eqn{eq:sprop}. The gauge transformation of the latter is given by \Eqn{eq:gtexpl}.
Clearly, the results of \Sec{sec:sym} directly apply to this case. Note that this includes the possibility of spontaneously broken gauge symmetry.

For what concerns the construction of gauge-invariant approximations to the 2PI effective action, the situation is more involved than in spinor QED, see \Sec{sec:gaugeinvapprox}, owing to the presence of both a derivative trilinear coupling and a quartic coupling in the classical action \eqn{appeq:SQED}. When involved in a given diagram of the 2PI expansion, the former acts on the scalar leg of (mixed) propagators attached to the vertex, giving rise to a nontrivial gauge transformation of the diagram. Similarly, the quartic coupling gives rise to an $A^\mu$-dependent trilinear vertex in the shifted action ${\cal S}[\varphi+\hat\varphi]$, which also leads to a nontrivial gauge transformation of any diagram in which it is involved.
In order to build gauge-invariant approximations, one has to identify gauge-invariant subsets of diagrams, where these nontrivial gauge transformations cancel. 

The general procedure described in \Sec{sec:gaugeinvapprox} teaches us to organize the diagrammatic expansion in terms of the classical three- and four-point vertices
\bea
 \lambda^{(3)}_{\mu}(x,y,z;A)&\equiv&\frac{\delta^3S[A,\phi,\phi^\dagger]}{\delta A^\mu(z)\delta\phi(x)\delta\phi^\dagger(y)}\nn
 &=&-ie\delta^{(4)}(x-z)\delta^{(4)}(y-z)\Big[D_{\mu}^x(A)^\dagger-D_{\mu}^y(A)\Big]
\eea
and
\bea
 \lambda^{(4)}_{\mu\nu}(x,y,z,t)&\equiv&\frac{\delta^4S[A,\phi,\phi^\dagger]}{\delta A^\nu(t)\delta A^\mu(z)\delta\phi(x)\delta\phi^\dagger(y)}\nn
 &=&2e^2g_{\mu\nu}\delta^{(4)}(x-z)\delta^{(4)}(y-z)\delta^{(4)}(z-t)\,,
\eea
where we have introduced the covariant derivative $D_\mu^x(A)\equiv\partial_\mu^x-ieA_\mu(x)$.
Consider then a given 2PI diagram involving the derivative, field-dependent three-point vertex $\lambda^{(3)}$. Clearly, the covariant derivatives act on the scalar leg of (mixed) propagators attached to the vertex, giving rise to a mere local phase factor under a gauge transformation. The phase factors
associated with the two scalar legs attached to the vertex cancel with each other, in a similar way as in QED, see \Eqn{eq:gaugecancel}. A similar cancellation of phase factors arises at each four-point vertex $\lambda^{(4)}$.

\end{document}